\documentclass[traditabstract]{aa}
\usepackage[pdftex]{graphicx}
\usepackage[varg]{txfonts}
\usepackage[english]{babel}
\usepackage{natbib}
\usepackage{pdfpages}
\usepackage[caption=false]{subfig}
\usepackage{comment}
\usepackage{amssymb}
\usepackage{url}

\begin{document}
\title{SUBARU and e-Merlin observations of NGC3718}
\subtitle{Diaries of an SMBH recoil?}

\author{K. Markakis\inst{1,2}, 
   	J. Dierkes\inst{5},
        A. Eckart\inst{1,2},
        S. Nishiyama\inst{4},
        S. Britzen\inst{2},
        M. Garc\'iía-Mar\'in\inst{1},
        M. Horrobin\inst{1},
        T. Muxlow\inst{3},
        \and
        J. A. Zensus\inst{2,1}
        }

\institute{I. Physikalisches Institut, Universit\"at zu K\"oln,
           Z\"ulpicher Str. 77, 50937 K\"oln, Germany \\
           \email{markakis@ph1.uni-koeln.de}
           \and
           Max-Planck-Institut f\"ur Radioastronomie,
           Auf dem H\"ugel 69, 53121 Bonn, Germany \\
           \and
           Jodrell Bank Centre for Astrophysics, School of Physics and Astronomy, 
           The University of Manchester, Oxford Road, Manchester M13 9P \\
           \and
           Miyagi University of Education, Sendai, Miyagi 980-0845, Japan\\
           \and
          G\"ottingen eResearch Alliance, State and University Library G\"ottingen, Papendiek 14, 37073 G\"ottingen\\
             }

\date{Received 29 September 2014 / Accepted Day-Month-Year}

\abstract{
NGC3718 is a LINER $L1.9$ galaxy, lying at a distance of about $\sim 17.4$ Mpc away from earth and its similarities with NGC5128 often award it the name ``northern Centaurus A''. The presence of a compact radio source with a candidate jet structure, a prominent dust lane, and a strongly warped molecular and atomic gas disk are indicative that NGC3718 has undergone some sort of a large scale gravitational interaction sometime in the recent past. This channeled gas towards the center, feeding the black hole and igniting the central engine. One proposed scenario involves an encounter with the close neighboring galaxy NGC3729, while other authors favor a merging event with mass ratio $\geq(3-4):1$, as the origin of NGC3718. 

We use high angular resolution ($\sim100$ mas) e-Merlin radio and SUBARU NIR ($\sim170$ mas) data, to take a detailed view of the processes taking place in its central region. In order to preserve some objectivity in our interpretation, we combine our results with literature values and findings from previous studies. Our NIR maps suggest, on one hand, that towards the stellar bulge there are no large scale absorption phenomena caused by the apparent dust lane and, on the other, that there is a significant (local) contribution from hot ($\sim1000$ K) dust to the nuclear NIR emission. The position where this takes place appears to be closer to the offset compact radio emission from our e-Merlin $6$ cm map, lying offset by $\sim4.25$ pc from the center of the underlying stellar bulge. The shape of the radio map suggests the presence of one (or possibly two, forming an X-shape) bipolar structure(s) $\sim1$ ($\sim0.6$) arcsec across, which combined with the balance between the gas and the stellar velocity dispersions and the presence of hard X-ray emission, point towards effects expected by AGN feedback. We also argue that NGC3718 has a ``core'' in its surface brightness profile, despite the fact that it is a gas-rich galaxy and we discuss its mixed photometric and spectroscopic characteristics. The latter combined with the observed spatial and radio offsets, the relative redshift between the broad and the narrow $H{\mathrm{\alpha}}$ line, the limited star formation activity and AGN feedback, strongly imply the existence of an SMBH recoil. Finally, we discuss a possible interpretation, that could naturally incorporate all these findings into one physically consistent picture.

}   
   
\keywords{Galaxies: kinematics and dynamics -- Galaxies: active -- Galaxies: evolution -- Galaxies: formation -- Galaxies: photometry}

\titlerunning{NGC3718 - inside the heart of a LLAGN}
\authorrunning{Markakis et al.}

\maketitle

\section{Introduction}
\label{sec:intro}
Knowledge of the circumnuclear activity in active galaxies is essential for understanding the fueling of the central engine. Our understanding of the underlying physical processes that contribute to the nuclear activity is, however, still far from being complete and conclusive. While on large scales ($\geq$ $\sim 3$ kpc) the picture is clearer, as large scale dynamical perturbations i.e. galaxy collisions, mergers \citep[e.g.][]{1972ApJ...178..623T} etc., are proposed as the mechanisms responsible for removing angular momentum from the gas, driving it towards the central region, the corresponding processes at smaller scales (sub-kpc) are not very well understood. Mechanisms ranging from nested bars \citep[e.g.][]{1989Natur.338...45S} and spirals \citep[e.g.][]{1999AJ....118.2646M} to warped nuclear disks \citep{2000ApJ...533..850S,2000ApJ...533..826S} and  $m=1$ instabilities \citep{1999ApJ...522..772K,2000A&A...363..869G}, have been proposed over the years in order to explain the smaller scale phenomena, the discussion, however, is still open. 

\begin{figure}[h]
\centering
\includegraphics[width=4.9cm]{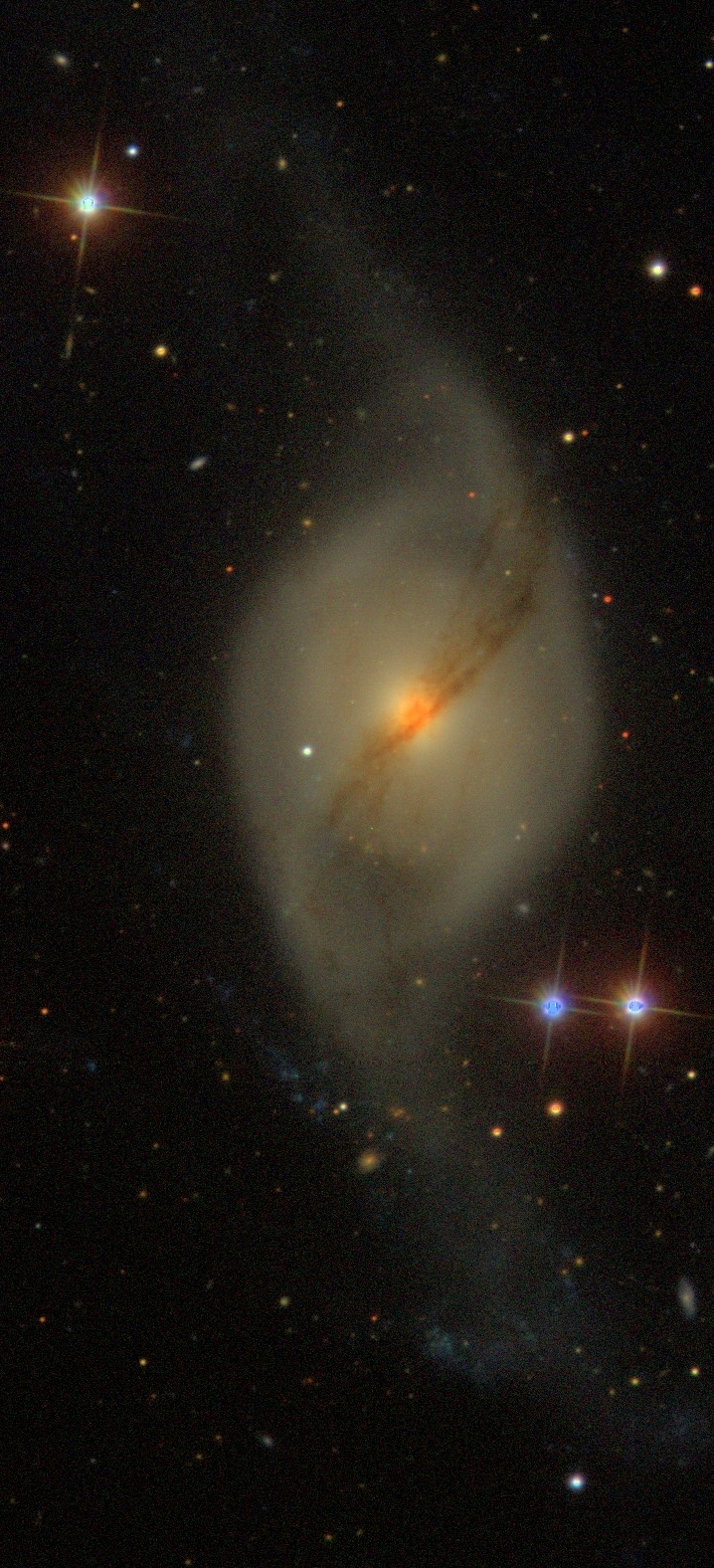}
\caption{$9.5\times 4.5$ arcmin SDSS $gri$ composite image of NGC3718.}
\label{fig:SDSS_comb}
\end{figure}

An ideal laboratory for trying to shed light on some of these important questions is NGC3718 (figure \ref{fig:SDSS_comb}). NGC3718 and its (supposed) companion NGC3729 belong to the loose Ursa Major group, with NGC3718 being one of the largest galaxies in the group. It is unclear, however, whether these two galaxies interact gravitationally or not and to what extent \citep{1996AJ....112.2471T}. \cite{2013MNRAS.429.2264K} find that both galaxies belong to the sub-group NGC3992 (named after the gravitationally dominant member), which is one of the most massive sub-groups of Ursa Major. They also note that, velocities within the NGC3992 sub-group do not show any visible correlation with distances, indicating a non-virialized system. Morphologically, NGC3718 is classified as an SB(s)a pec by \cite{1991rc3..book.....D}, mainly due to the prominent -''spiral arm'' ending- dust lane which was considered to be indicative of the presence of a bar. \cite{1958MeLu2.136....1H} classified NGC3718 as an S0p, whereas other authors \citep[e.g.][]{1990AJ....100.1489W,1994A&A...291...57R}, classify it as a polar ring galaxy, confusing our understanding of its true morphology.

The distinguishing features of NGC3718 are the prominent dust lane, which runs across the entire stellar bulge, and its strongly warped molecular and atomic gas disk. Several authors have extensively studied the gas dynamics of NGC3718. \cite{1985A&A...142..273S} studied the HI dynamics and found that the atomic gas distribution forms a 3-dimensional warped structure, which could be described by tilted but concentric rings, orbiting from nearly edge-on at smaller radii, to nearly face-on at larger radii. \cite{2004A&A...415...27P} and \cite{2005A&A...442..479K} studied the molecular gas distribution using $CO(1\rightarrow0)$, $CO(2\rightarrow1)$ and $HCN(1\rightarrow0)$ as tracers. They successfully fit tilted rings on NGC3718 and found that the molecular gas motion, generally, follows that of the HI gas but the warp continues down to scales of $\sim250$ pc. \cite{2009AJ....137.3976S} re-mapped the HI distribution using higher resolution VLA data and confirmed (though with slightly different parameters) the tilted ring models of the aforementioned studies. They show that the outer gas orbits extend to $\sim 35-42$ kpc and they estimate its age at $\sim2-3$ Gyr, whereas the inner gas orbits are nearly polar and still under formation. Finally, they do not see any HI gas in the plane of rotation of the stellar disk and they argue in favor of the classification of NGC3718 as a polar ring galaxy, invoking differential precession in order to account for the warp. 

Some of these features are also present in NGC5128, host of the famous Centaurus A radio source. A similar dust lane \citep{1979AJ.....84..284D} and a warp in the gas disk \citep[e.g.][]{1997A&A...322..419W,1996ApJ...473..810S}, along with the presence of radio emission (though more dominant) from the nucleus, are common properties which often lead to the characterization of NGC3718 as ``the Northern Centaurus A''.

NGC3718 is also one of the NUGA sources, a survey aimed at the study of nearby low-luminosity active galactic nuclei (LLAGN) \citep{2003ASPC..290..423G}. Spectroscopically, it is classified as a LINER type $L1.9$ galaxy \citep{1997ApJS..112..391H}. A weak broad $H{\mathrm{\alpha}}$ emission component with $FWHM_{H{\mathrm{\alpha}},Broad} \approx 2350$ km s$^{-1}$ is detected, originating from the nucleus. Additionally, the presence of strong $[O I]\lambda\lambda 6300 \AA$ with $FWHM_{[O I]} \approx 570$ km s$^{-1}$ is indicative of a hidden AGN \citep{1985ApJS...57..503F}. \cite{2007A&A...464..553K} detect a candidate jet structure in a $18$ cm Merlin radio map, lying NW of the nucleus and stretching to $\sim0.5$ arcsec. They also measure the bolometric luminosity of $\sim10^{41}$ ergs s$^{-1}$ which implies a sub-Eddington system.

\cite{2002A&A...388..407C} treat ARP214 (an alternative name for NGC3718) as an advanced merger remnant, while \cite{2002A&A...393L..89J} discuss its mixed characteristics. Photometrically, it shows an exponential light profile (like a spiral galaxy) but kinematically, it is mainly supported by pressure from random motion of stars (like an elliptical galaxy), as indicated by the ${V_{C} \over \sigma}\sim 0.5-1$ within the inner few kpc. \cite{2005A&A...437...69B} simulated galaxy mergers with mixed photometric and kinematic characteristics through N-body simulations and they find that such objects could result from mergers with mass ratios  $\geq(3-4.5):1$.

This paper is organized as follows: In section \ref{sec:observations} we present the data and data reduction, in section \ref{data_processing} we describe the pre-processing and alignment procedures, in section \ref{sec:nir maps} we present the NIR maps, we attempt a light decomposition and we present the 6 cm e-Merlin radio map, in section \ref{sec:bh_relations} we discuss the scaling relations, the classification and the observed mixed characteristics of NGC3718, in section \ref{sec:EM_sign} we present additional observational evidence regarding the presence or not of an SMBH recoil and in section \ref{sec:discussion} we try to put all the pieces together in order to get a physically consistent picture. Finally, in appendices \ref{sec:app_A}, \ref{sec:app_B} and \ref{sec:app_C}, we provide additional information for the various arguments we present in this paper.

Throughout this paper we adopt the following: In all images North is up and East is left. The cosmology values used are $\Omega_{M} = 0.27$, $\Omega_{\Lambda} = 0.73$, $H_{0} = 67.8$ km s$^{-1}$, with a redshift of $z = 0.003927$ corrected to the reference frame defined by the 3K CMB\footnote{\label{NED}\url{http://ned.ipac.caltech.edu/cgi-bin/objsearch?search_type=Obj_id&objid=26880&objname=1&img_stamp=YES&hconst=67.8&omegam=0.27&omegav=0.73&corr_z=1}}. These suggest a distance modulus $\mu = 31.2$ and a distance $D = 17.4$ Mpc for NGC3718. The cosmological scale for the adopted cosmology at this distance is $84$ $pc \over arcsec$. We also use the ``Mass to Light'' ratio relationship from \cite{2003ApJS..149..289B} for all the conversions from light to mass. The general form of their $M \over L$ ratio in the NIR and optical (B-V) colors is:

\begin{equation}
\log\left(\frac{M_*/M_\odot}{L/L_{\odot,\lambda}}\right) = a_\lambda + b_\lambda (B-V).
\label{eq:mlratio}
\end{equation}

with $a_{\lambda}$ and $b_{\lambda}$ being $a_J=-0.261,a_H=-0.209,a_K=-0.206,b_J=0.433,b_H=0.210$, and $b_K=0.135$. For NGC3718 this value is ${M \over L} \sim 0.78$, using $M_{V}^{NGC3718} = -20.73$ mag and $M_{B}^{NGC3718} = -20.01$ mag, magnitudes corrected for extinction and K-corrections, taken from NED (footnote \ref{NED}).

\section{Observation and data reduction}
\label{sec:observations}

\subsection{SUBARU data}
\label{sec:SUB_data}
The data set used for the analysis of NGC3718 consists of, AO assisted near infrared (NIR) data in $J$ ($5$), $H$ ($11$) and $K_S$ ($12$) bands (number of images) respectively, taken on the 17th of May 2012, with the SUBARU telescope at Mauna Kea, Hawaii, using the HiCIAO \citep{2010SPIE.7735E..30S} instrument, operating in the Direct Imaging Mode. The individual frame exposure time for all bands is $t^{JHK_S}_{exp} = 60$ sec, with the total duration of the observations being $\sim1h~40m$. During this time, the total variation of airmass is $\sim0.082$. The AO188 AO system is used \citep{2010SPIE.7736E..0NH}. It is equipped with a 188-element wavefront curvature sensor with photon counting APD modules and a 188 element bimorph mirror, installed at the IR Nasmyth platform of the Subaru telescope, which for this dataset operates in self reference mode on the core of NGC3718. As a result, the angular resolution of the data is $\sim170$ mas. The $2048\times 2048$ pixels$^2$ Hawaii-IIRG HgCdTe detector provides a pixel scale of $0.010$ $arcsec \over pixel$, with a FoV (Field of view) of $20 \times 20$ arcsec$^2$. 

\begin{figure}[h]
\centering
\includegraphics[width=4cm]{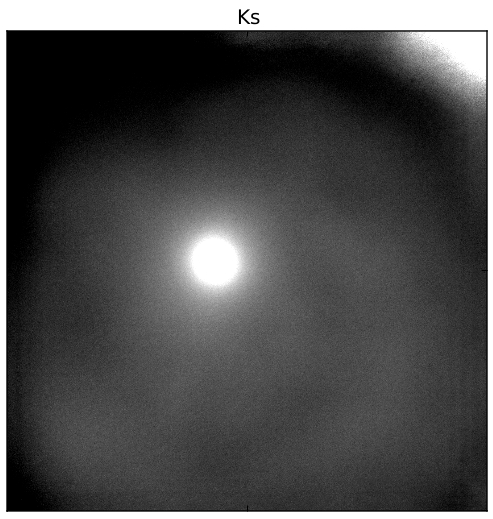}
\includegraphics[width=4cm]{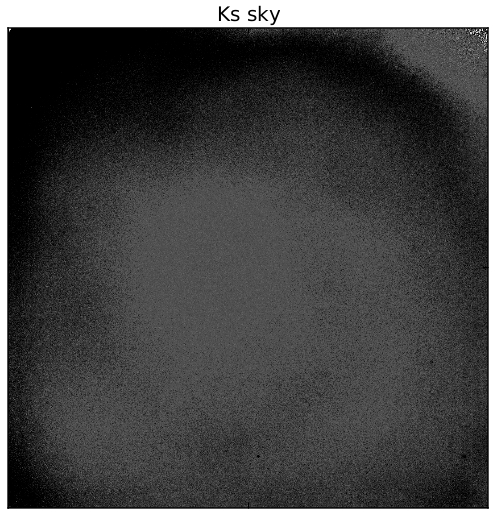}
\caption{$K_S$ band before sky subtraction (left) and the sky frame extracted from the $K_S$ image (right). Similar sky frames were created and subtracted from the $J$ and $H$ bands, as well.}
\label{fig:sky_frame}
\end{figure}

\begin{figure}[h]
\centering
\includegraphics[width=4cm]{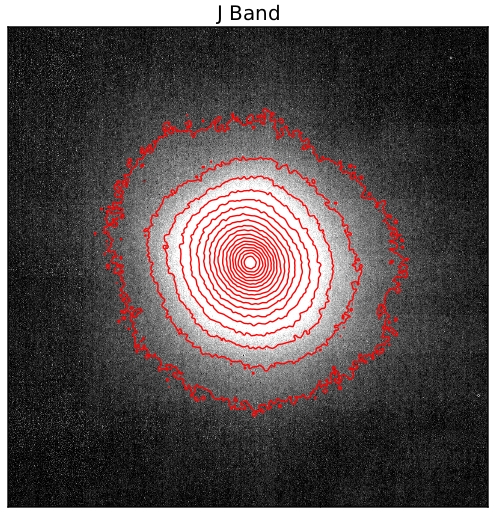}
\includegraphics[width=4cm]{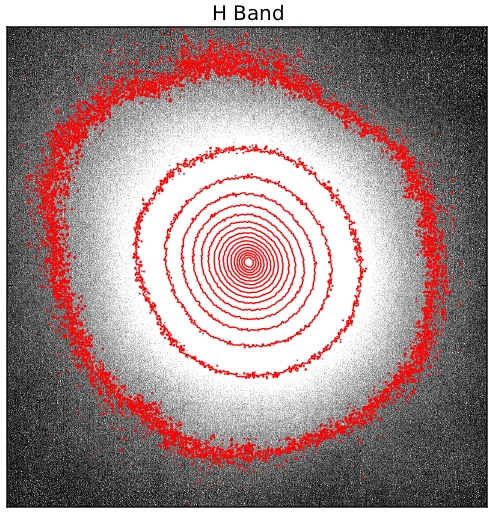}
\includegraphics[width=4cm]{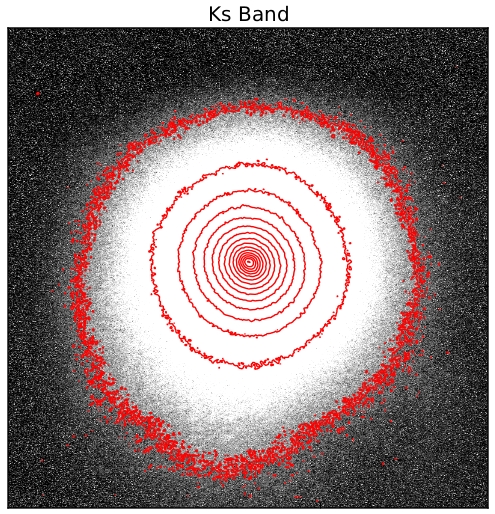}
\caption{$J$(top left), $H$(top right) and $K_S$(bottom middle) ready-for-science images.}
\label{fig:data}
\end{figure}

No reduction package was available for HiCIAO, so a pipeline was developed from scratch, in order to correct for the high frequency 32-strip artifact noise, introduced by the 32 readout channels of the detector. All images were dithered and have undergone the usual bad pixel correction, flat-fielding (dome-flat), alignment (based on ellipse fitting) and median coadd treatment. The final images allow us to generate sky frames for each band from the data themselves, by clipping them near the modal background value, and adding a high frequency layer on top\footnote{Perhaps a more correct term would be a ``background'' subtraction, but in the NIR the majority of the background contamination refers to the sky component. We create our ``sky'' frame(s) by keeping values lower than the mode of the image(s), which should be a good approximation of the sky contamination. On top of that, we add a layer obtained by applying a high pass filter on the image(s), in order to account for the pixel-to-pixel variations as well.}(figure \ref{fig:sky_frame}). The data with the sky subtracted are shown in figure \ref{fig:data}.

\subsection{e-Merlin data}
\label{sec:merlin_data}
The e-MERLIN synthesis telescope is a seven element interferometer, with baselines of up to $217$ km and connected by a new optical fiber network to Jodrell Bank Observatory near Manchester, UK. An inhomogeneous array, e-MERLIN is comprised of the $76$ m Lovell telescope, a $32$ m dish at Cambridge, and the following $25$ m antennas: Mark II, Knockin, Defford, Pickmere and Darnhall. 

The data in this work were taken during the commissioning phase of e-MERLIN with only the five $25$ m dishes, resulting in a primary beam of $10'$, a maximum baseline of $133.7$ km (between Pickmere and Defford) and a minimum baseline of $11.2$ km (between Pickmere and Mark II). The final, fully expanded, array will have a bandwidth of $2$ GHz, providing more than 10 times the continuum sensitivity as the original MERLIN. NGC3718 was observed at $5$ GHz for $10$ hrs by e-MERLIN on the $3$rd of August $2011$, with $4$ $128$ MHz sub-bands of $512$ spectral channels each, yielding a total bandwidth of $512$ MHz. The observations did not include baselines to Cambridge. The final angular resolution we obtain is $\sim100$ mas. 

Data were reduced and analyzed using the National Radio Astronomy Observatory's (NRAO) Astronomical Image Processing System (AIPS). Data were initially edited with SPFLG and IBLED, averaged to $64$ channels per intermediate frequency (IF) channel and concatenated with DBCON before further editing was conducted. FRING was used to derive delay and rate corrections for the calibrator sources, while CALIB was used to derive time-dependent phase and then amplitude and phase solutions. Flux calibration was performed using short observations of 3C286 at the beginning and end of each run and the flux density scale was calculated with SETJY \citep{2013ApJS..204...19P}. The flux density for each IF was then reduced by $4\%$, in order to account for the resolution of the e-MERLIN shortest spacing (see the MERLIN User Guide). Bandpass and phase calibration was performed using the bright point sources. Absolute calibration is expected to be accurate to about $\sim10\%$ for commissioning data.

The quoted positional errors, associated with phase transfer errors from phase reference to target source (for a typical 3$^o$ separation), are $\sim$1~mas. This assumes 6-stations including Cambridge and online L-Band link corrections. The phase reference source used in our e-Merlin observations is J1146+5356 and we have referenced our observations to the coordinates given in the VLBA calibrator list, namely $\alpha_{2000}$=11:46:44.204328 and $\delta_{2000}$=+53:56:43.08356. This places J1146+5356 about 2.28$^o$ away from NGC3718. Taking into account the lower resolution of these data without Cambridge, and making some allowance for known small phase corrections for the L-Band link transmission delays around our observational epoch, we conservatively estimate the phase transfer errors to be around $\sim$4~mas. The phase reference source position from the VLBA Calibrator list quoted formal positional uncertainty errors of 0.35~mas and 1.17~mas in R.A. and Dec., respectively. The formal measurement error from the peak of the e-Merlin image will be 100/2Q~mas with a restoring beam of 100~mas and a peak signal to noise in the image of Q which is better than 60 in our case. Hence the total estimate of the positional uncertainties result in a value of $\sim$4.1~mas.

\section{Data processing}
\label{data_processing}

\subsection{Centering}
\label{sec:centering}

\begin{figure}[h]
\centering
\includegraphics[width=\columnwidth]{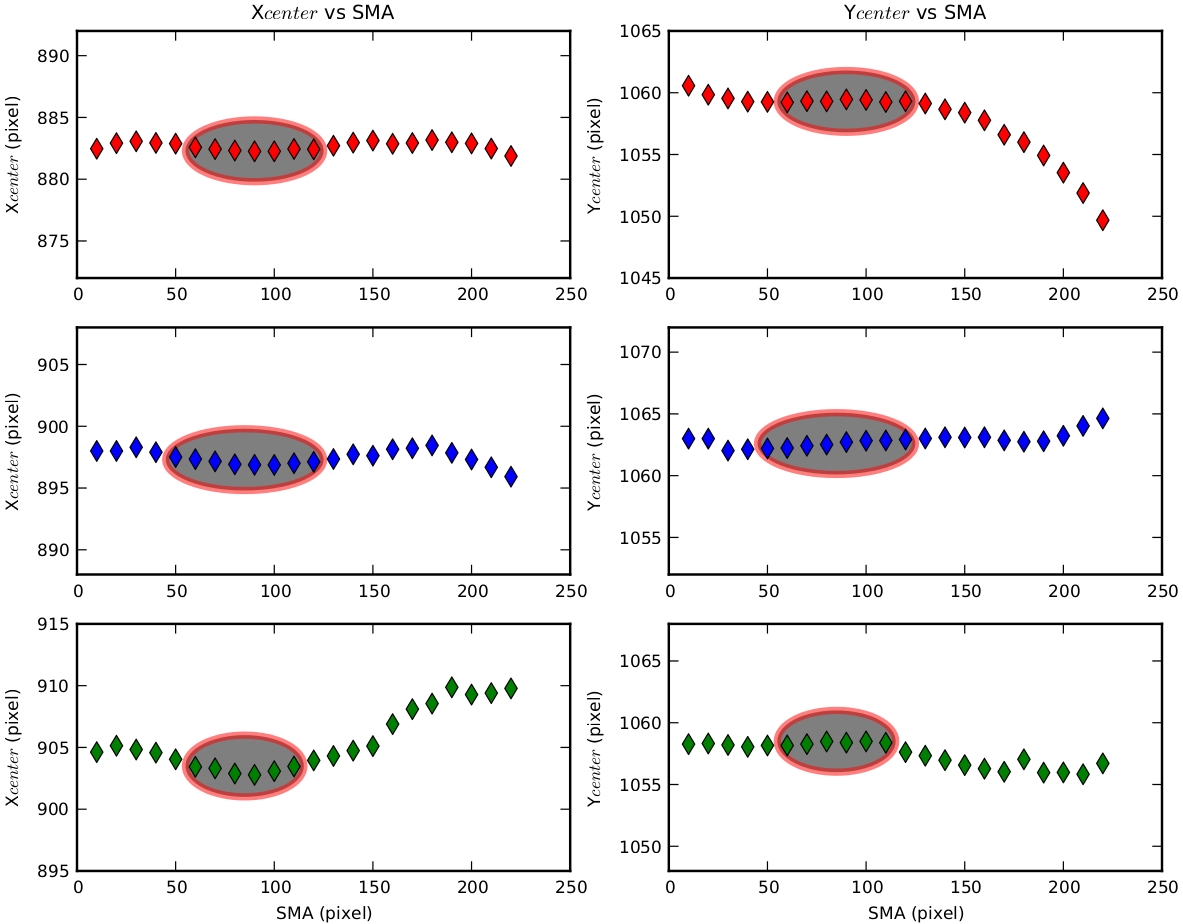}
\caption{X (left column) and Y (right column) central coordinate of each fitted ellipse versus its SMA, for the $K_S$(red/top), $H$(blue/middle) and $J$(green/bottom) bands. The shaded ellipses indicate the \emph{most prominent stable values} used to derive the first order approximation photocenters.}
\label{fig:init_plot}
\end{figure}

The first step of our SUBARU data analysis, is to align the $J$,$H$ and $K_S$ images. In the absence of secondary sources in our FoV, we choose an alignment based on \emph{ellipses-on-isophote contours} fitting. For this purpose we use IRAF's\footnote{IRAF is distributed by the National Optical Astronomy Observatories, which are operated by the Association of Universities for Research in Astronomy, Inc., under cooperative agreement with the National Science Foundation.} task \emph{ellipse}, which fits ellipses to iso-intensity contours of a galaxy's light distribution. We plot the $X_{center}$ and $Y_{center}$ coordinates indicated by each fitted ellipse versus its S(emi) M(ajor) A(xis), as shown in figure \ref{fig:init_plot}. 

Apart from the $X_{center}$ coordinate of the $J$ band and the $Y_{center}$ coordinate of the $K_S$ band, the rest of the curves behave relatively normally\footnote{In the absence of asymmetries in the light distribution, the plots of figure \ref{fig:init_plot} are expected to be close to horizontal lines i.e. all the fitted ellipses should point towards a common center.}. The behavior of the $J$ band, can be attributed to the lower, with respect to the $H$ and $K_S$ bands, total flux, $F_{J}$ ($F_{K_S}\sim3F_{J} $ and $F_{H}\sim2.5F_{J}$ respectively). This is partly due to the lower integration time ($t_J=300$ sec, $t_H=660$ sec and $t_{K_S}=720$ sec), but it mainly indicates that the stellar flux is of the order of the one expected, from an evolved stellar population of K/M type stars\footnote{\label{old_stars}The $V-K$ color of $NGC3718$ is $\sim3$ (NED), suggesting a dominant stellar population with $T_{eff}\sim4000$ K, consistent with K/M type stars \citep{1980ApJ...235..126R}.}. The lower flux results in more noisy isophote contours and, consequently, in higher uncertainties. The dust lane of NGC3718 does not seem to affect the light distribution in the NIR to a large extent, mainly because it lies far away enough (see section \ref{sec:nir maps} and Appendix \ref{sec:app_A}). Moreover, the dominant component, both in terms of structural size and illumination, should be the stellar bulge of the galaxy. The absence of large scale contour deformations, and the general contours shape\footnote{The NIR ellipticities range between $\sim0.09-0.12$ in $J$ band, $\sim0.07-0.09$ in $H$ band and $\sim0.03-0.07$ in $Ks$ band, in the radial interval between $\sim0.5-1.5$ arcsec.}, indicate a largely relaxed stellar component in all three NIR bands. 

We therefore consider, that the average of the \emph{most prominent stable values} of the ($X_{center}$ , $Y_{center}$) coordinate pairs (points within the shaded ellipses of figure \ref{fig:init_plot}), should act as a very good first order approximation of the photocenter (i.e. center of the stellar bulge) for each band. The uncertainty of each ($\bar{X}_{center}$ , $\bar{Y}_{center}$) pair, is considered to be the quadratic addition of the \emph{standard deviation of the most prominent stable values} used for each band with their average fitting uncertainties ($\bar{dX}_{center}$ , $\bar{dY}_{center}$) as computed by the \emph{ellipse} task itself. The suggested photocenters and their corresponded uncertainties, are displayed in table \ref{tab:init_fit}.

\begin{table}[h]
\caption{First order approximation photocenters.}
\centering
\label{tab:init_fit}
\begin{tabular*}{\columnwidth}{@{\extracolsep{\fill}}ccccc}\hline \hline
Filter & $\bar{X}_{center}$ & $\bar{Y}_{center}$ & $\bar{dX}^{total}_{center}$ & $\bar{dY}^{total}_{center}$ \\
 & (Pixel) & (Pixel) & (mas) & (mas) \\\hline

$J$ & 903.16 & 1058.36 & 3.4 & 2.2\\
$H$ & 897.10 & 1062.59 & 2.5 & 3.0\\
$K_S$ &882.39 & 1059.33 & 1.5 & 1.3\\

\hline
\end{tabular*}
\tablefoot{
First order estimation of $\bar{X}_{center}$ $\&$ $\bar{Y}_{center}$ and the uncertainties involved in mas, as suggested by ellipse fitting on the $J$, $H$ and $K_S$ bands. $1$ pixel equals $10$ mas.\\
}
\end{table}

\subsection{Pivot and Subtract}
\label{sec:Pivot and Subtract}

Having a first order approximation of the center of the stellar bulge, the next question is: ``What could be the cause of the deviation in the central coordinates of the $K_S$ band?''. A clean bulge's light distribution should have a nice symmetrical bell shape structure (e.g. a Sersic profile), as are all spherical/elliptical shapes that are projected on a 2-D surface. This should lead to co-centered isophotes and, therefore, to co-centered fitted ellipses.

So we consider the following scheme: A theoretical perfectly symmetrical 3-D structure should look the same under any rotation around the perpendicular to the X-Y plane axis, if we have a spherical shape, or at least under $n \times \pi$ rotation, with n being an integer, if we have an elliptical shape. So if one was to subtract a 2-D projection of such structure (a bell shape curve) from a rotated (around its center) copy of itself, one should receive 0 as a result. Following this scheme for a perfectly symmetrical, isolated and undistorted bulge, if one rotates and subtracts its 2-D projection from itself, one should not receive any obvious residual pattern, apart from random noise. In the case of NGC3718 we have an elliptically projected bulge, so this scheme is valid only under $n \times \pi$ rotation around the first order approximation photocenters, derived from the curves of figure \ref{fig:init_plot}, with $n=1,2,3,...n$.

\begin{figure}[h]
\centering
\includegraphics[width=4cm]{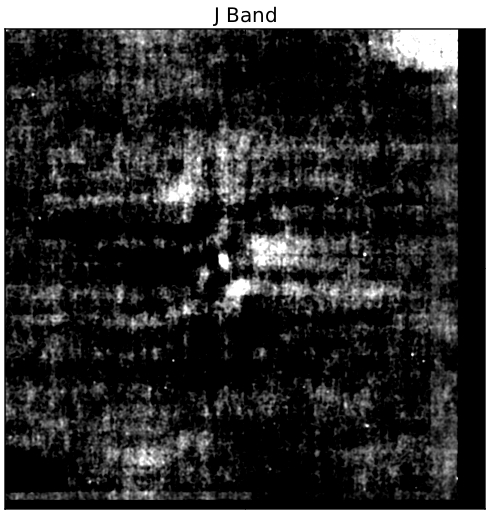}
\includegraphics[width=4cm]{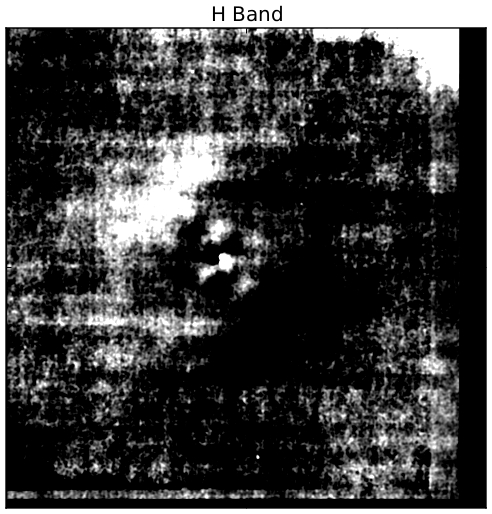}
\includegraphics[width=4cm]{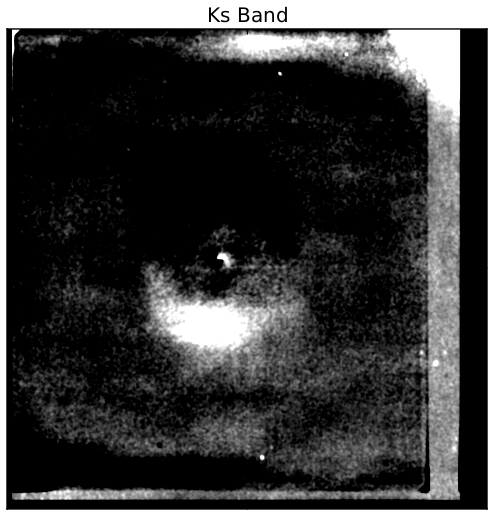}
\caption{$J$(top left), $H$(top right) and $K_S$(bottom middle), smoothed with a 10 pixel Gaussian kernel, residual images from the pivot and subtract operation. The residual mean peak fluxes, expressed as a percentage of $NGC3718$'s mean peak fluxes, are $\sim4.5\%$($J$), $\sim3.0\%$ ($H$) and $\sim3.4\%$ ($K_S$) respectively.}
\label{fig:init_resid}
\end{figure}

The pivot ($\pi$ rotation) and subtract operation reveals strong residuals, especially in the $K_S$ and $H$ bands (figure \ref{fig:init_resid}), whereas in the $J$ band the residuals are considerably more noisy.

\subsection{Evaluating the centering}
\label{sec:evaluating the centering process}

Despite the fact that the uncertainties implied are of the order of a few mas (see table \ref{tab:init_fit}), we test the robustness of our initial center estimations, by shifting the pivot point in all bands by ($X^{init}_{center}\pm1,Y^{init}_{center}\pm1$) and ($X^{init}_{center}\pm0.5,Y^{init}_{center}\pm0.5$) and repeating the pivot and subtract operation. In this way, we test a series of \emph{sixteen} alternative pivot points, arranged in a rectangular (2 pixels in side) pattern around the initial center estimations. If the mean central ($300\times300$ pixels$^2$) residual produced by a candidate ($X^{i}_{center},Y^{i}_{center}$) is smaller than the mean residual of the ($X^{init}_{center},Y^{init}_{center}$), then it is adopted as a better center estimation and the process is iterated, until the minimum mean residual is reached. 

\begin{figure}[h]
\centering
\includegraphics[width=4cm]{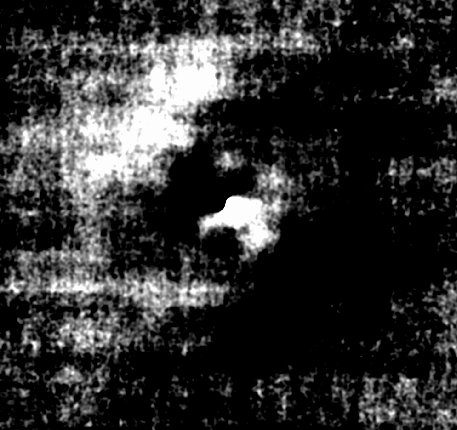}
\includegraphics[width=4cm]{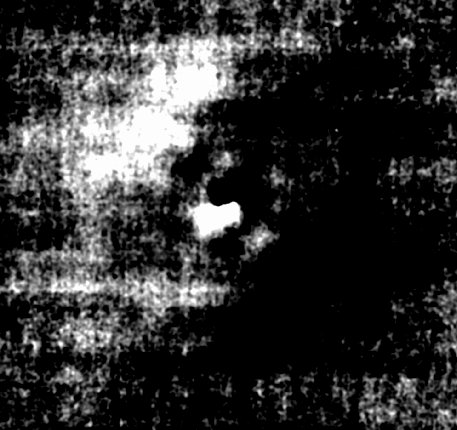}\\
\includegraphics[width=5cm]{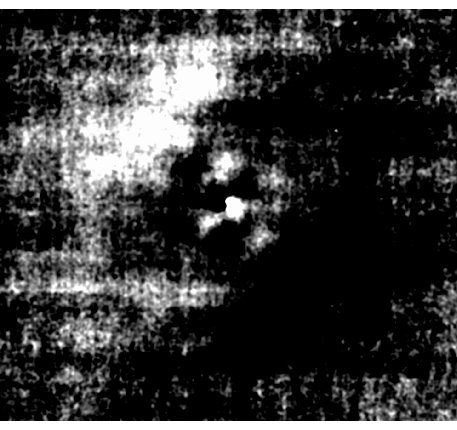}\\
\includegraphics[width=4cm]{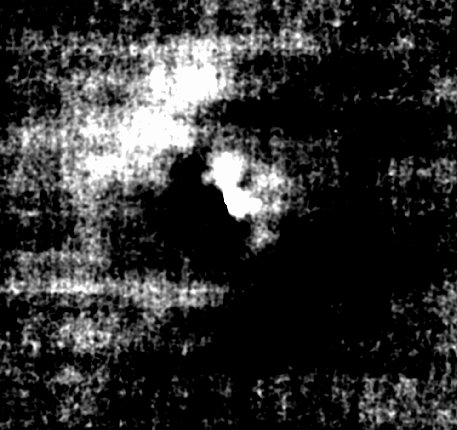}
\includegraphics[width=4cm]{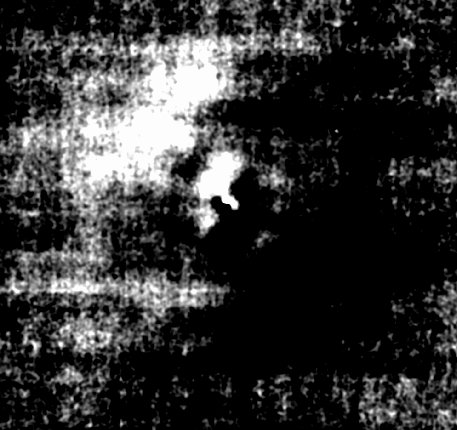}
\caption{Examples of the $H$ band smoothed residual images after a rotation around incorrect pivot points, by ($X^{init}_{center}-0.5,Y^{init}_{center}+0.5$) (upper left), ($X^{init}_{center}+0.5,Y^{init}_{center}+0.5$) (upper right), ($X^{init}_{center},Y^{init}_{center}$) (center), ($X^{init}_{center}-0.5,Y^{init}_{center}-0.5$) (lower left) and ($X^{init}_{center}+0.5,Y^{init}_{center}-0.5$) (lower right).}
\label{fig:init_resid_false}
\end{figure}

Our initial center estimations proved to be quite accurate, at least within $\pm0.5$ pixels ($\pm5$ mas) per coordinate, since these are the pivot points that produce the minimum mean residuals. This is slightly larger than the statistical uncertainties of table \ref{tab:init_fit}, and we adopt them, as a more conservative approach. An example of the  above process for the $H$ band, is shown in figure \ref{fig:init_resid_false}. In general, the residuals produced by the alternative pivot points are a lot less well defined, especially in the central region. The brighter central features that appear are indicative of the, deliberately, incorrect pivot points used. The extended light excess towards the NE from the center, is largely unaffected, so the larger scale asymmetries can still be traced and removed. 

\begin{figure}[h]
\centering
\includegraphics[width=4cm]{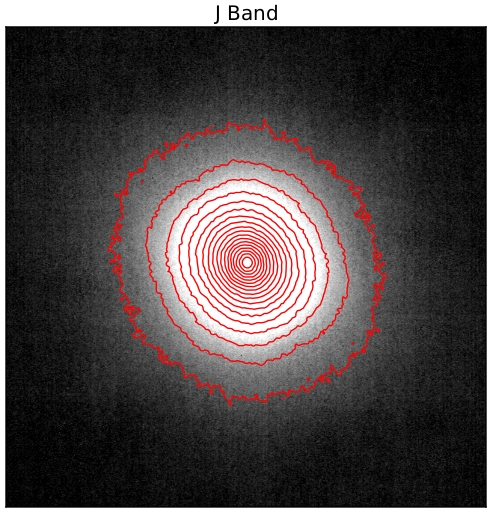}
\includegraphics[width=4cm]{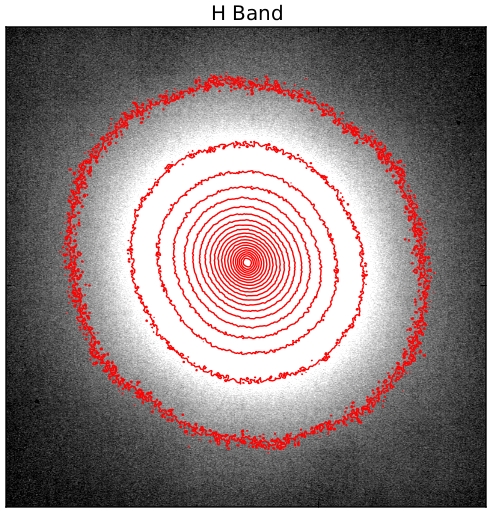}
\includegraphics[width=4cm]{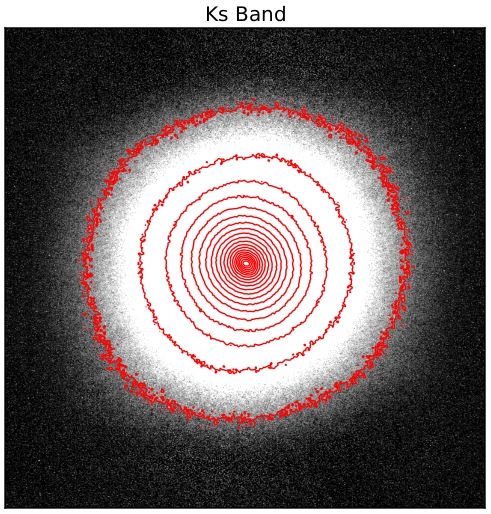}
\caption{$J$(top left), $H$(top right) and $K_S$(bottom middle) symmetrical light images after the removal of the residuals.}
\label{fig:bulges}
\end{figure}

We remove the residuals of figure \ref{fig:init_resid} from the images of figure \ref{fig:data}, in order to clear the light distribution from all the asymmetries, as shown in figure \ref{fig:bulges}. The symmetrical light images in all bands are clearly more round and generally, better defined.

We note that the residual images represent asymmetries between diametrically opposite parts of the light distribution. In the presence of inclination and/or tilt, the further we move from the center, the line-of-sight light path (and therefore the brightness) difference between diametrically opposite parts of the galaxy become increasingly larger, so that, any large scale asymmetries can not be simply treated as light over-densities with a physical meaning. As we approach the center however, these effects become increasingly smaller, so that, in small scales the observed asymmetries should mainly represent true light (and therefore mass) over-densities\footnote{As we approach the center, the brightness difference becomes increasingly smaller. When the residuals are minimal, with respect to neighboring pivot points, they should represent, to a first order, actual light density variations.}. The latter suggests that we can safely use the residual maps and the symmetrical light images for extracting physically robust quantities, as long as we restrict ourselves in small scales\footnote{Following Ockham's razor, however, throughout this paper we try to restrict the use of either of these to minimum, for the sake of credibility.}.

\subsection{Angular resolution matching $\&$ flux calibration}
\label{sec:matching angular resolution and flux calibration}

Having an accurate center estimation, we align our images with respect to the $J$ band. We match the angular resolution of our images, by convolving the better seeings ($H$ and $K_S$) to the worst one ($J$), using the FWHMs as measured from our PSF-reference stars in each band. The Gaussian kernel used, equals to the \emph{quadratic difference} of the smaller from the larger FWHM, as follows. 

\begin{equation}
\delta_{FWHM}(H) = \sqrt[]{[\mathrm{FWHM(J)^2} - \mathrm{FWHM(H)^2}]} \approx 9.8
\end{equation}
\begin{equation}
\delta_{FWHM}(K_S) = \sqrt[]{[\mathrm{FWHM(J)^2} - \mathrm{FWHM(K_S)^2}]} \approx 10.3
\end{equation} 

\noindent
Flux calibration was performed, using our standard stars, to derive Zero Points for each band according to: 
\begin{equation}
m_i = \mathrm{ZP_i} - 2.5 \log([\mathrm{counts_i}/\mathrm{Exp.Time_i}])
\end{equation}

\subsection{Coordinate uncertainties}
\label{sec:uncertainties}

In order to translate the relative into an absolute coordinate system, the following method is used: We align our SUBARU $J$ band with a re-sampled and re-scaled\footnote{We magnify a $51\times51$ pixels$^2$ (FoV $20\times20$ arcsec$^2$) part of the SDSS $z$ band centered on the nucleus of NGC3718, by a factor of $\sim40$, equal to the scale ratio of the SDSS and SUBARU data (namely $\sim0.4$ and $\sim0.01$ $arcsec \over pixel$, respectively). The new, $2040\times2040$ pixel$^2$, SDSS $z$ band replica is then aligned with our SUBARU $J$($H$ and $K_S$, trimmed to $2040\times2040$ pixel$^2$) band(s), following the method described in section \ref{sec:centering}.} version of the SDSS $z$ band image, in order to (astrometrically) calibrate it. We choose the SDSS $z$ band, because its central wavelength ($0.9134 \mu$m) is the nearest to our $J$ band ($1.220 \mu$m), so the centroids of similar, old stellar populations are expected to coincide in the NIR, (see Appendix \ref{sec:app_B} and footnote \ref{old_stars}), and also because the SDSS is the most accurate (in astrometric terms) survey currently available \citep{2003AJ....125.1559P}. The total intrinsic SDSS astrometric uncertainties were calculated following the instructions described in the SDSS online documentation. The resulted values for our entire uncertainty budget, are shown in table \ref{tab:uncert}. 

\begin{table}[h]
\caption{Astrometric uncertainties and coordinates.}
\centering
\label{tab:uncert}
\begin{tabular*}{\columnwidth}{@{\extracolsep{\fill}}cccccc} \hline \hline
 uncertainties:& SDSS & $z$-$J$ & Cent./Shift & e-Merlin & Total \\
 & (mas) & (mas) & (mas) & (mas) & (mas)\\\hline

$J$ band\\\hline

$\delta{(R.A.)}$&42.5 & 10.8 & 5.0 & 4.1 & 44.3\\
$\delta{(Dec.)}$&41 & 11.0 & 5.0 & 4.1 & 42.9\\\hline\hline

$H$ band  \\\hline

$\delta{(R.A.)}$&42.5 & 10.8 & 7.1 & 4.1 & 44.6\\
$\delta{(Dec.)}$&41 & 11.0 & 7.1 & 4.1 & 43.2\\\hline\hline

$K_S$ band \\\hline

$\delta{(R.A.)}$&42.5 & 10.8 & 7.1 & 4.1 & 44.6\\
$\delta{(Dec.)}$&41 & 11.0 & 7.1 & 4.1 & 43.2\\
\end{tabular*}

\begin{tabular*}{\columnwidth}{@{\extracolsep{\fill}}ccccc} \hline \hline
& R.A. (Bulge)& Dec. (Bulge)\\
&11:32:34.8515 $\pm$ 0.0051 & +53:04:04.475 $\pm$ 0.044\\\hline\hline
& R.A. (Red blob)& Dec. (Red blob)\\
&11:32:34.8550 $\pm$ 0.0051 & +53:04:04.512 $\pm$ 0.044\\\hline\hline
& R.A. (Radio)& Dec. (Radio)\\
&11:32:34.8497 $\pm$ 0.0005\tablefootmark{*} & +53:04:04.527 $\pm$ 0.004\\\hline\hline
\end{tabular*}
\tablefoot{Upper table: Total astrometric uncertainties involved in our results interpretation for each band. Columns: 1) Total intrinsic SDSS astrometric uncertainties, 2) $z$-$J$ bands matching uncertainties during the SUBARU astrometric calibration, 3) centering and/or shifting uncertainties during $J$,$H$,$K_S$ centering and the $J$-$H$,$J$-$K_S$ alignments, 4) intrinsic radio map astrometric uncertainties, 5) quadratic addition of all the above. Lower table: From top to bottom, positions of the peak flux of a) the stellar bulge of NGC3718, b) the offset red blob and c) the offset e-Merlin $6$ cm radio emission.\\
\tablefoottext{*}{This is the position of the peak of the radio emission of NGC3718, using the position of J1146+5356 quoted in section \ref{sec:merlin_data}. Assuming the position used in earlier e-MERLIN observations, our R.A. and Dec. measurements namely, 11:32:34.8534 $\pm$ 0.0005 and +53:04:04.523 $\pm$ 0.004, are, within the uncertainties, in very good agreement with the position published by \cite{2007A&A...464..553K}.}
}
\end{table}

\section{NIR color maps}
\label{sec:nir maps}

\begin{figure}[h]
\centering
\includegraphics[width=\columnwidth]{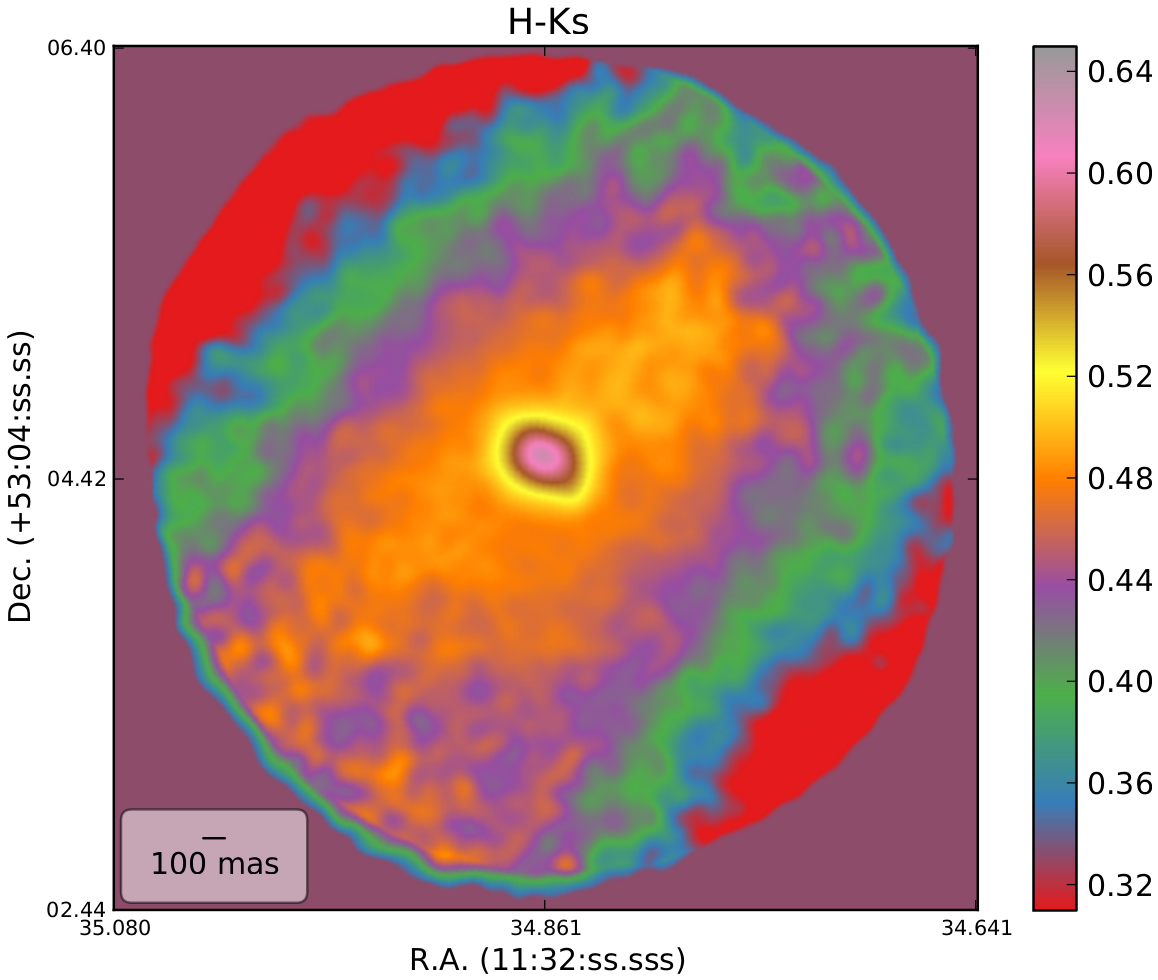}
\includegraphics[width=\columnwidth]{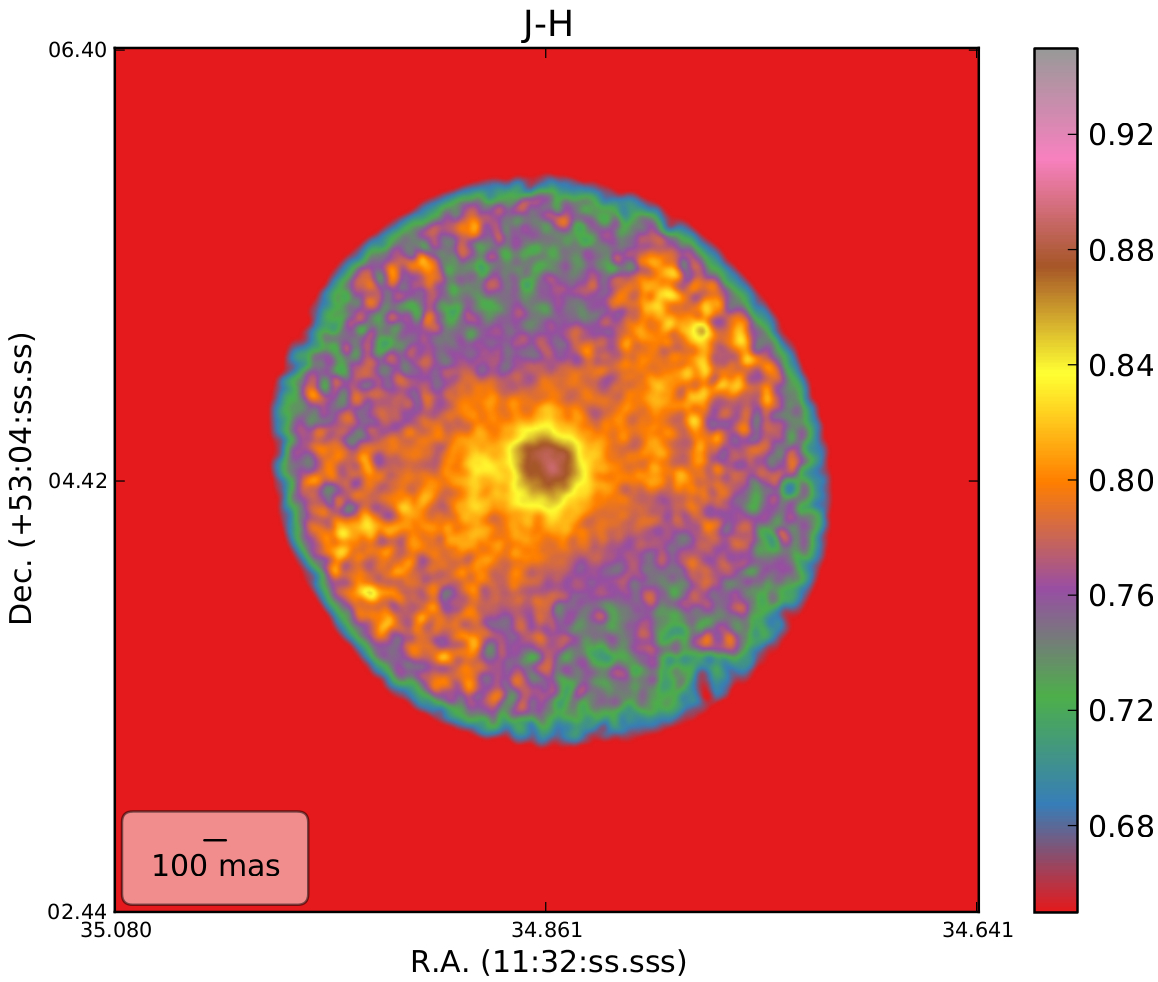}
\caption{$H-K_S$(top), $J-H$(bottom) NIR color maps.}
\label{fig:NIR_maps_1}
\end{figure}

Having aligned and calibrated our NIR images, we proceed by producing the $J-H$ and $H-K_S$ NIR color maps. NIR color maps are an important tool as they provide vital information about the color distribution of a galaxy, such as whether reddening affects the colors and whether light comes purely from an ``ordinary'' stellar component or if it contains a contribution from a nuclear component as well \citep{1985MNRAS.214..429G}. The colors of ``ordinary'' galaxies are, $J-H=0.78$ and $H-K=0.22$ \citep{1984MNRAS.211..461G}. Light captured by the $J$ band is produced mainly from stars, whereas in $K_S$, light can come from both stars and hot dust. A black body with a temperature below the sublimation temperature of dust ($\sim1500$ K), will radiate at $\lambda \gtrsim 1.9 \mu$m, mainly affecting the $K_S$ band but, potentially, also the $H$ band ($\lambda_{central} = 1.63 \mu$m) to some extent. In this context, $J-H$ is mostly indicative of extinction, as dust is already evaporated, and is acting as a scattering source, while  $H-K_S$, is indicative for hot dust emission. The NIR color maps are shown in figure \ref{fig:NIR_maps_1}.

Our $J-H$ map is fairly symmetric. An approximately elliptical red excess can be seen surrounding the nuclear region extending from SE to NW. This could be interpreted as local extinction, since there is no apparent \emph{large scale constant color gradient}, which would indicate that the dust lane (lying SW and \emph{outside} the borders of the light distribution, see Appendix \ref{sec:app_A}) extends all the way to the projection of the center of the bulge. The $H-K_S$ map also shows a similar, but better defined, elliptical red excess around the center (sharing the same orientation with the excess in the $J-H$ map), indicative for the presence of hot dust emission. 

\begin{figure}[h]
\centering
\includegraphics[width=\columnwidth]{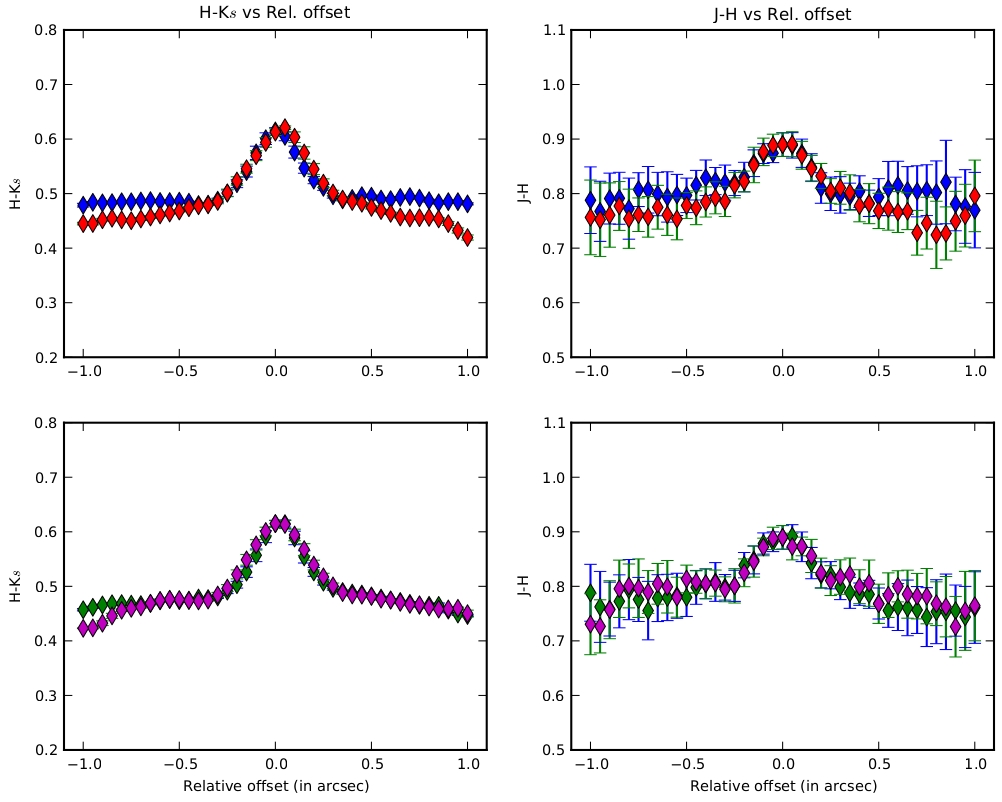}
\caption{Radial profiles of the $H-K_S$ (left column) and $J-H$ (right column) NIR color maps, derived from two sets of perpendicular directions measured counterclockwise from the North. These are, $120\,^{\circ}/210\,^{\circ}$ (blue/red points respectively) in the upper row and $170\,^{\circ}/260\,^{\circ}$ (green/magenta points respectively) in the lower row. Negative$\rightarrow$positive relative offset reflects a S(E,W)$\rightarrow$N(W,E) direction.}
\label{fig:rad_plot}
\end{figure}

The overall symmetry of the NIR color maps can be more clearly seen on the radial profile plots of figure \ref{fig:rad_plot}. In the four different directions of measurement (see caption of figure \ref{fig:rad_plot}), the plots do not show any sign of a \emph{large scale constant color gradient}, which would appear as an asymmetry with respect to the center and it would indicate the presence of foreground dust\footnote{A non-uniform, progressively thinner, dust extension northward of the dust lane, would translate into progressively excessive dust towards the S/SW in comparison to the N/NE. This would lead to excessive reddening, resulting in an asymmetry in the figure \ref{fig:rad_plot} plots, in the form of: progressively more dust $\rightarrow$ progressively increasing reddening $\rightarrow$ progressively fainter $J$($H$) $\rightarrow$ progressively larger $J-H$ ($H-K_S$) towards the S/SW. A uniformly distributed foreground dust extension, however, could be present, but in this case the reddening should be uniform as well, a fact that would not affect (at least) the (qualitative) interpretation of our results.}. The only asymmetry that can be seen is between the blue and red profiles outside of $\sim0.5$ arcsec. This is due to the fact that the directions of measurement are perpendicular to each other and cross the center along the ``major'' and the ``minor'' axis of the red elliptical excess respectively. In any case, both of the components are highly symmetrical with respect to the center. 

\begin{figure}[h]
\centering
\includegraphics[width=\columnwidth]{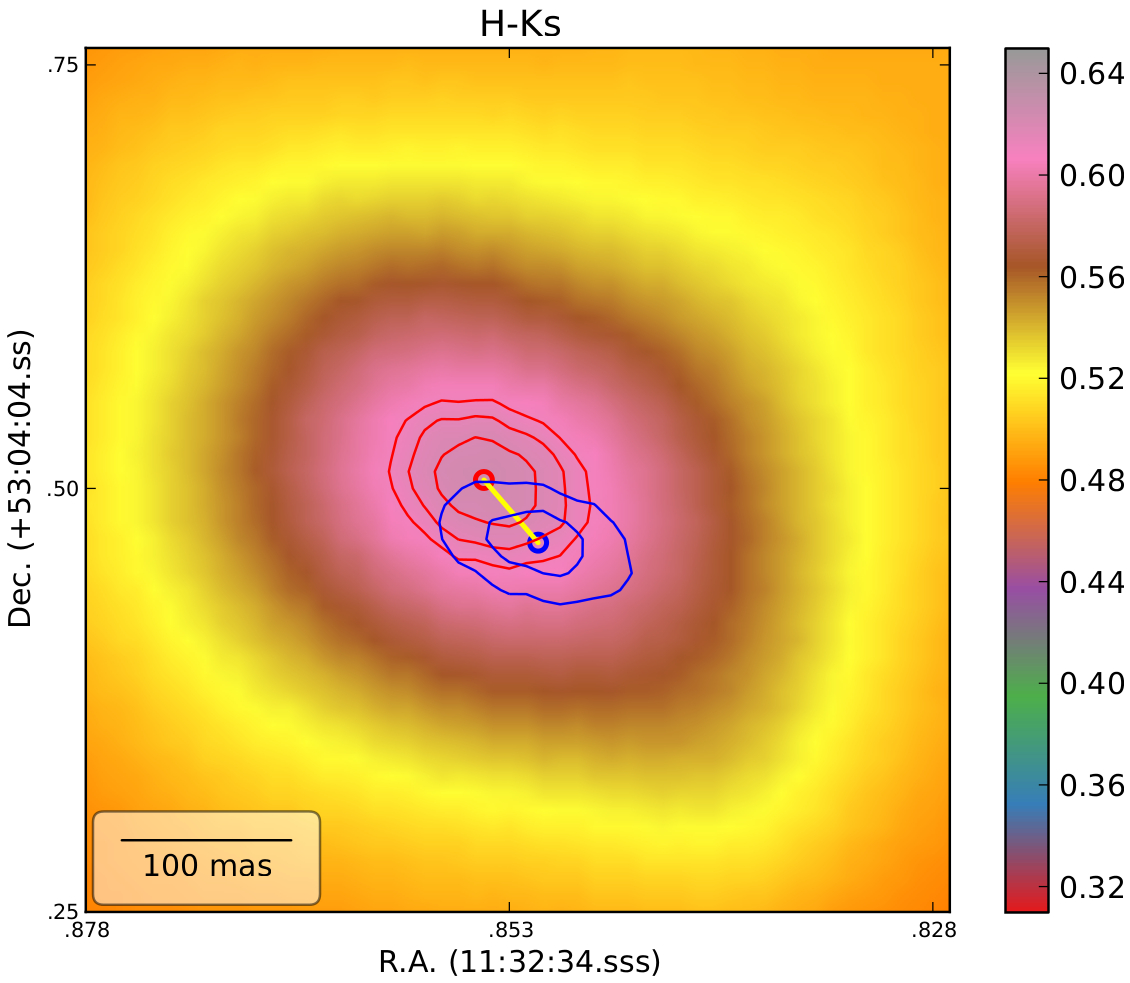}
\includegraphics[width=\columnwidth]{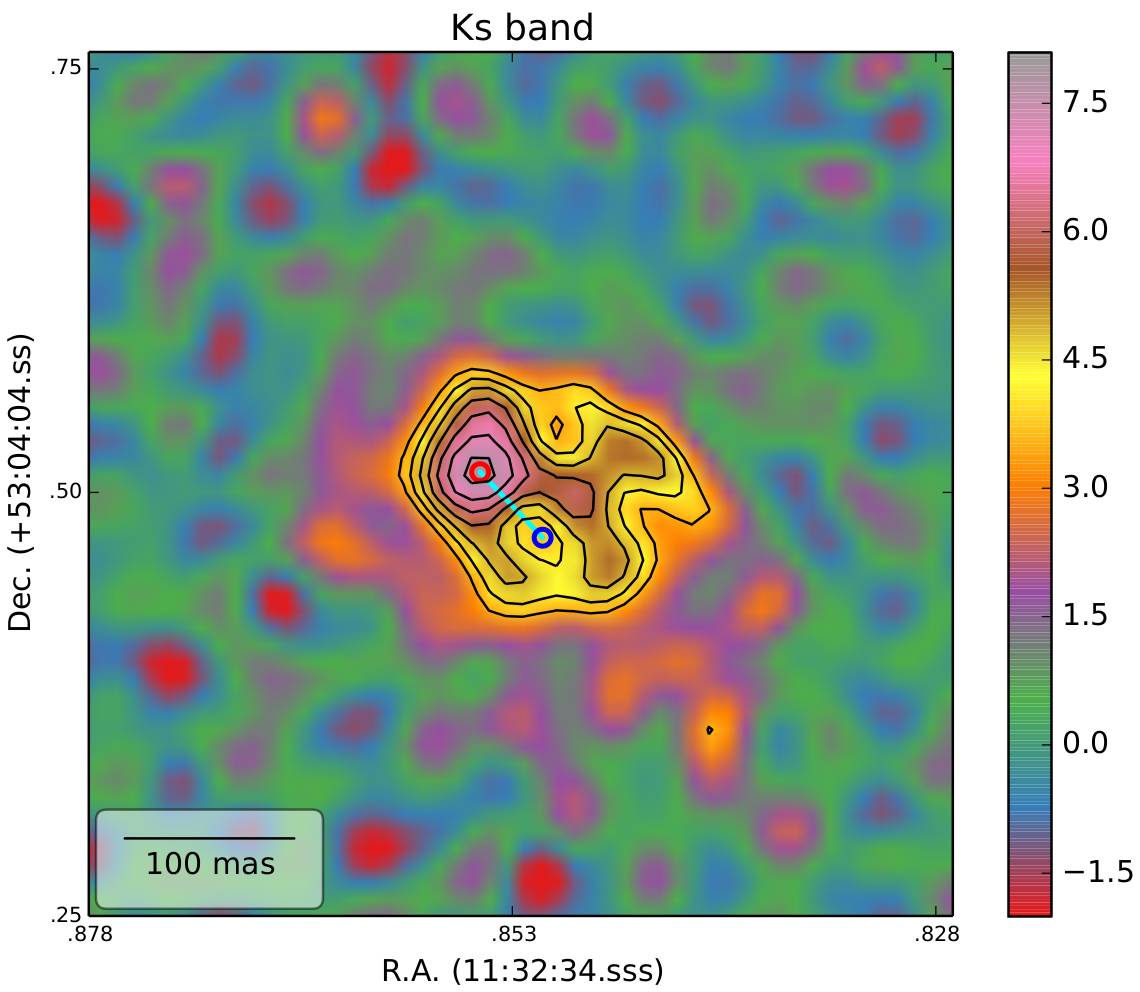}
\caption{Top: $H-K_S$ closeup. The red blob contours (red) are offset from the symmetrical light image's peak contours (i.e. the center of the stellar bulge, in blue). Bottom: $K_S$ band high pass filtered image closeup. The peak flux is offset from the photocenter of the $K_S$ band. In both images, the red circle indicates the position of the peak value of the offset red blob, while the blue circle indicates the position of the center of the stellar bulge, as derived from our contour analysis. Their radii indicate the positional uncertainty of each position, which is $\sim7$ mas. The yellow(top)/cyan(bottom) line is the length of the offset, namely $\sim 50$ mas.}
\label{fig:NIR_maps_2}
\end{figure}

The most interesting feature of the $H-K_S$ map, however, is that the center of the red blob (i.e. the peak of the $H-K_S$ map) lies slightly offset from the center of the stellar bulge, by $\sim50$ mas. For this to be more clearly seen, we produce the symmetrical $H-K_S$ map (using the $H$ and $K_S$ images of figure \ref{fig:bulges}) and we over-plot its peak contours (i.e. the center of the stellar bulge), alongside the peak contours of our normal $H-K_S$ map, in figure \ref{fig:NIR_maps_2}. We also manage to extract the same offset length, along the same orientation, by applying a high pass filter directly on the reduced $K_S$ image, a result ``free'' from any geometric transformation uncertainties. The corresponding central astronomical positions for the stellar bulge and the offset red blob, respectively, are shown in table \ref{tab:uncert}.

\subsection{Light decomposition}
\label{sec:light_decomp}

\begin{figure}[h]
\centering
\includegraphics[width=\columnwidth]{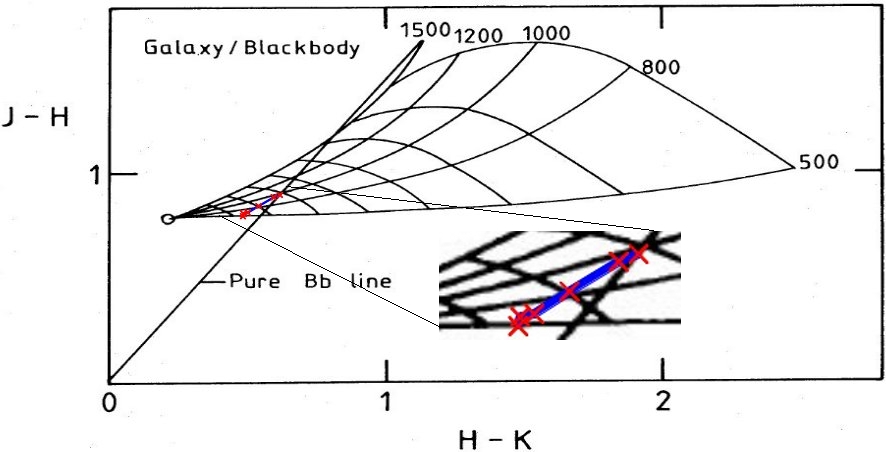}
\caption{ \cite{1985MNRAS.214..429G} two color diagram. Moving away from the ``ordinary'' galaxy colors (black circle at $J-H=0.78$ and $H-K=0.22$) and along the horizontal lines in the loci of mixtures, there is an increasing percentage ($10\%$ per vertical line) of contribution from hot dust emission, while moving along the vertical lines, the temperature of this dust component increases ($T\sim 500-1500$ K).}
\label{fig:glass}
\end{figure}

Following \cite{1985MNRAS.214..429G}, we use their figure 7 (see caption of figure \ref{fig:glass} for an explanation), of theoretical loci for mixtures of ``ordinary'' galaxy colors and blackbodies of various temperatures constrained on the color-color diagram of their sample, in order to decompose the light of NGC3718. An azimuthal average, binned in $5$ equally spaced ($\sim 50$ mas step) radius bins, of the innermost $0.5$ arcsec of the nucleus of NGC3718, can be seen in (the zoomed-in section of) figure \ref{fig:glass}. A decline, from a $50\%$ stellar - $50\%$ dust (at $T_{100} \sim 1000$ K) light mixture in the innermost $\sim 100$ mas, to a $60\%$ stellar - $40\%$ dust (at $T_{200} \sim 800$ K) in the innermost $\sim 200$ mas, which drops even further to a $65\%$ stellar - $35\%$ dust (at $T_{300-500} \sim 700-500$ K) in the innermost $\sim 300-500$ mas, can be clearly seen in this plot. This indicates the presence of significant contribution from hot dust in the central region of NGC3718, suggesting an environment ideal for SMBH accretion.

\subsection{Over-plotting the e-Merlin radio map}
\label{sec:merlin}

\begin{figure}[h]
\centering
\includegraphics[width=\columnwidth]{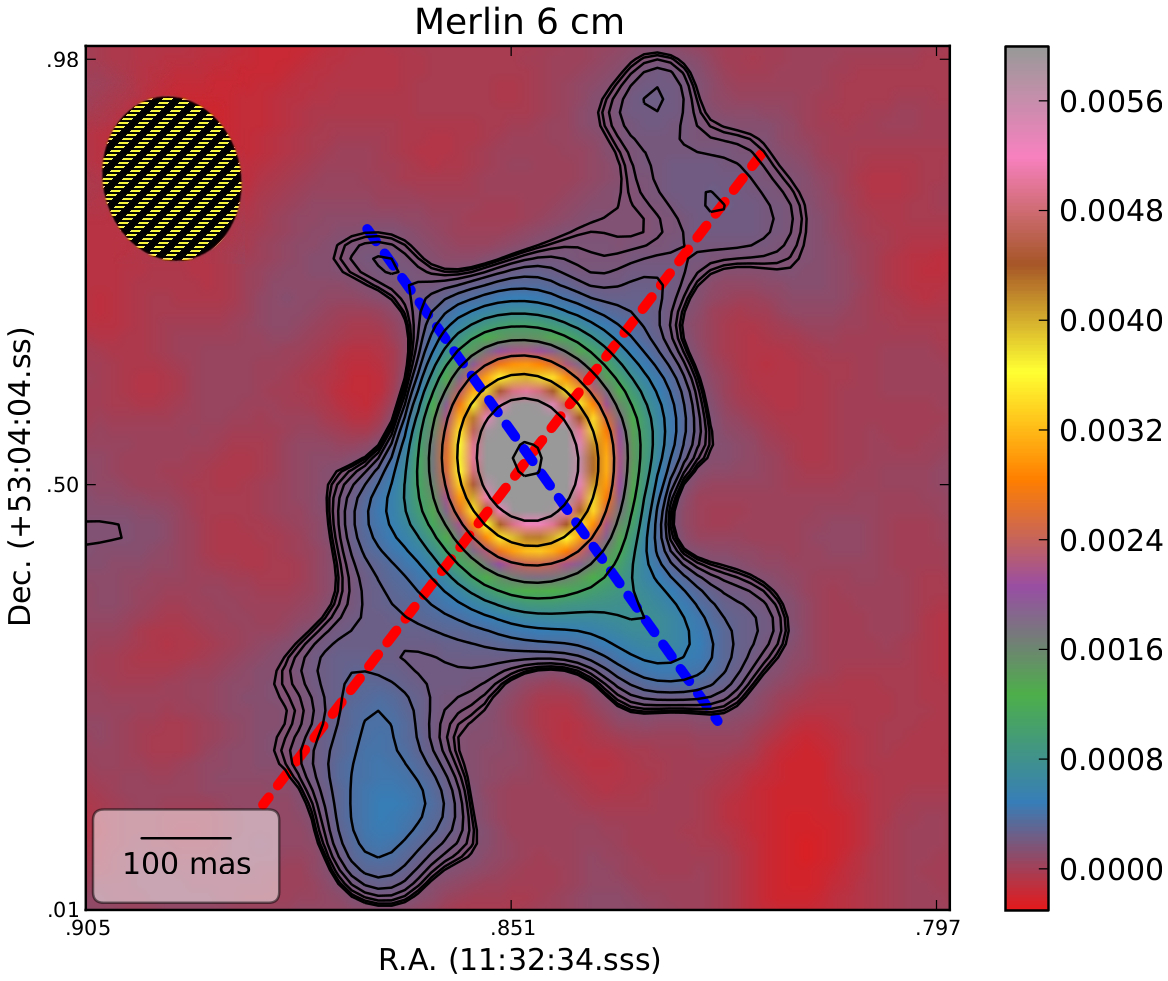}
\caption{e-Merlin $6$ cm radio map. The outer contours are set to $\sim4\sigma$ above the background where the flux is $\sim2$ orders of magnitude less than the peak flux. The third inner contour is $\sim40-50\%$ of the peak flux. The red and blue dashed lines, represent the position angle of the SE-NW and the NE-SW candidate small scale bipolar extended structures, respectively. The beam is shown in the upper left corner.}
\label{fig:emerlin}
\end{figure}

The high angular resolution ($\sim100$ mas) $6$ cm radio e-Merlin map, with a pointing accuracy of $\sim4$ mas (see section \ref{sec:merlin_data}), is  shown in figure \ref{fig:emerlin}. The outer contours are set to $\sim4\sigma$ sigma above the background, so we consider the flux density enclosed within these contours to be a true detection of extended radio emission, at least, in the position angle shown by the red dashed line in figure \ref{fig:emerlin}.

\begin{figure}[h]
\centering
\includegraphics[width=\columnwidth]{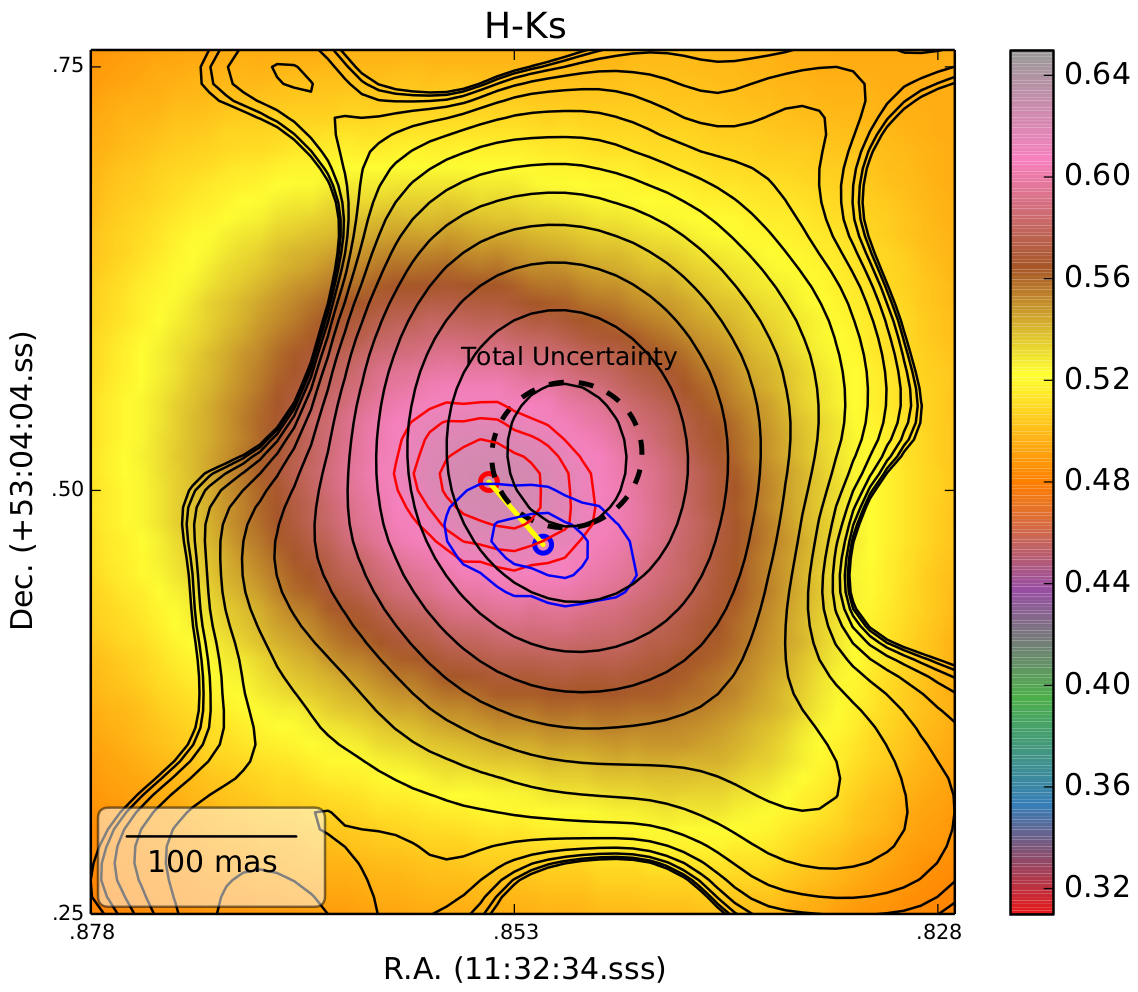}
\caption{Same as the upper image of figure \ref{fig:NIR_maps_2} but with the radio contours over-plotted (in the position quoted in table \ref{tab:uncert}. The same image with the equivalent to the \cite{2007A&A...464..553K} position, is shown in Appendix \ref{sec:app_C}, figure \ref{fig:H_Ks_emerlin_old_calib}). The blue and red circle radii, indicate the positional uncertainty of each position, namely $\sim7$ mas. The black dashed ellipse indicates the total coordinates uncertainty of table \ref{tab:uncert}.}
\label{fig:H_Ks_emerlin}
\end{figure}

Over-plotting the radio map on our $H-K_S$ NIR map, reveals that, the radio emission appears to originate from a position closer to the offset red blob than to the center of the stellar bulge (figure \ref{fig:H_Ks_emerlin}). This is yet another very interesting result, because it could indicate that the SMBH does not lie exactly at the center of the bulge, but it is offset by $\sim50$ mas. This, at a distance of $\sim$ $17.4$ Mpc, yields a \emph{projected} offset of $\sim4.25$ pc. The presence of radio emission is expected, since NGC3718 is a LLAGN, so the closer proximity of the radio emission to the offset red blob, implies that the SMBH accretes matter normally, but it does so slightly away from the center of the stellar bulge. The coordinate uncertainties however (see table \ref{tab:uncert}), are of the order of the offset, though little smaller. This suggests, that a potential coincidence of the radio emission with the offset red blob, is reasonably likely to be true. The astronomical position of the peak radio flux, is shown in table \ref{tab:uncert}.

Spatial offsets, potentially coincident with AGN offsets, are predicted by current theory. A possible interpretation could be that this is the case of a \emph{SMBH recoil} \citep[e.g.][]{2012AdAst2012E..14K,2011MNRAS.414.3656S}. In short, in the case of a merger, the SMBHs present in the centers of the merging galaxies will form a binary \citep{1980Natur.287..307B} which will eventually coalesce due to anisotropic emission of gravitational waves \citep[e.g.][]{1973ApJ...183..657B}. We need, however, more evidence to favor such a view of NGC3718, and will resume the discussion on this subject later on this paper.

We also confirm the presence of the $\sim0.5$ arcsec NW candidate jet component, that \cite{2007A&A...464..553K} see in their Merlin $18$ cm map. Moreover, we detect a diametrically opposite placed $\sim0.5$ arcsec extension towards SE. This could be considered as a possible counterpart of that jet, since, both lie along the same orientation and are of the same length. Finally, a less extended but brighter structure towards SW\footnote{In the position angle of the blue dashed line, which could be identified as the westward candidate jet component, \cite{2007A&A...464..553K} see in their EVN $6$ cm map. The same structure, aligns also with the jet-like structure seen at arcmin scales by the VLA at 1.49GHz in C/D-array configuration by \cite{1987ApJS...65..485C}.} and, a fainter tail-like one towards NE, both of $\sim0.3$ arcsec in length and sharing the same orientation, could be, potentially, indicative of a small scale X-shape radio structure, often associated with ``spin-flip'' processes in SMBH recoils \citep[e.g.][]{2012ApJ...746..176L} or with the two jet emitting members of a close SMBH binary \citep{2007MNRAS.374.1085L}.

Regarding the $6$ cm e-Merlin radio flux of NGC3718, we measure a peak flux of $8.85 \pm 0.07$ mJy and an integrated flux of $9.97 \pm 0.08$ mJy, both of which are larger than the values published by \cite{2007A&A...464..553K}, namely, $5.3 \pm 0.1$ mJy and $6.1 \pm 0.3$ mJy, respectively. This reflects an increase in peak flux of $\sim 67\%$ and in integrated flux of $\sim 63\%$, indicating that NGC3718 is a variable radio source, a picture consistent with an accreting SMBH.

\section{Scaling relations $\&$ classification}
\label{sec:bh_relations}

\subsection{To ``bulge'' or to ``pseudo-bulge''?}
\label{sec:bulge}

The most famous scaling relations between a host galaxy and its SMBH, are the $M_\mathrm{BH} - L$ and $M_\mathrm{BH} - \sigma$, relating the luminosity and the stellar velocity dispersion of the bulge, respectively, with the mass of the SMBH. 

We can estimate the absolute magnitude $M_{K_S,bulge}$ of NGC3718 from our data. Given the fact that we do not perform any decomposition into different components (disk, bulge etc.), the (contaminated) bulge magnitude we derive, should provide an \emph{upper limit} for the estimation of the $M_\mathrm{BH}$. We perform aperture photometry on our $K_S$ band images (figures \ref{fig:data} and \ref{fig:bulges}) of NGC3718, using three different aperture sizes, from $3-5\sigma$ above the background, with an average aperture of $r_{aper.} \sim 3$ arcsec\footnote{The average difference of the flux measurements on the images of figure \ref{fig:data} from those of figure \ref{fig:bulges}, is $\sim0.1$ mag which we consider negligible.}. We average the flux of those measurements and we use a (rather large) $0.2$ mag assumed\footnote{To account for the statistical deviation of the measurements, for systematic uncertainties in the calibration process etc.} error. The $K_S$ band absolute magnitude of NGC3718 is: 

\centerline{$M_{K_S,bulge} = -21.1 \pm 0.2$ mag~~.}

For the velocity dispersion however, we will need spectroscopic information. Luckily, NGC3718 is a very well studied object. \cite{1997ApJS..112..315H}, in their Palomar AGN spectroscopic survey, were able to extract several useful parameters, among those, the $FWHM[N II] = 371$ km s$^{-1}$ line. Assuming a Gaussian distribution and using $FWHM = 2\sqrt[]{(2ln(2))} \sigma \approx 2.355 \sigma$, we can get an estimation for the velocity dispersion, of $\sigma \approx 157$ km s$^{-1}$. However, $[N II]\lambda\lambda 6583 \AA$ is considered to be a better tracer for the ionized gas motion. A number of authors have suggested that the gas may rotate at a different speed than the stars (e.g. \cite{2001MNRAS.323..188P}, \cite{1986ApJ...305..136C}). \cite{2009ApJ...699..638H} combined gas and stellar velocity dispersion measurements and they conclude that $\sigma_*$ and $\sigma_g$ correlate well. \cite{2009ApJS..183....1H} calculate a velocity dispersion for NGC3718 of $\sigma = 158.1 \pm 9.6$ km s$^{-1}$, using the $Ca+Fe$ absorption feature at $6495\AA$, value which we adopt. This suggests a $\sigma_{gas} \over \sigma_{\star}$ $\sim 1$ for NGC3718. The author argues that ``\emph{...as the gas derives principally from mass loss from bulge stars, its kinematics should generally track the kinematics of the stars. But because the gas is collisional and experiences hydrodynamical drag against the surrounding hot medium, we expect it to be kinematically slightly colder than the stars. In the absence of additional energy input from other sources, we anticipate $\sigma_{gas} \over \sigma_{\star}$ $\lesssim 1$, as observed. As additional energy is injected into the system, for example from activation of the central black hole, the gas gains energy, to the point that $\sigma_{gas}$ approaches or even overtakes $\sigma_{\star}$}''. This picture is consistent with the fact that NGC3718 is indeed an LLAGN. This could be considered as an indirect indication that, NGC3718's central engine provides enough AGN feedback, to heat up the gas and cause this strong agreement between $\sigma_{gas}$ and $\sigma_{\star}$, information which will be used later on. 

Having an estimation of both of these quantities, we use the $M_\mathrm{BH} - M_{K_S,bulge}$ and $M_\mathrm{BH} - \sigma$ relations from \cite{2013ARA&A..51..511K}, namely:

\begin{equation}
\log{M_\mathrm{BH} \over 10{^9}M_{\odot}} = -(0.265 \pm 0.050) - (0.488 \pm 0.033)(M_{K_S,bulge} + 24.21)
\label{eq:M_L}
\end{equation}

\begin{equation}
\log{M_\mathrm{BH} \over 10{^9}M_{\odot}} = -(0.509 \pm 0.049) + (4.384 \pm 0.287)\log\left({\sigma \over 200 km s^{-1}}\right)
\label{eq:M_sigma}
\end{equation}

\noindent
Equation \ref{eq:M_L} suggests an SMBH mass:

\centerline{$M_{BH}^{M_{BH}-M_{K_S,bulge}} = 1.65^{+1.64}_{-0.82}\times 10^7 M_{\odot}$~~,}

\noindent
whereas equation \ref{eq:M_sigma} suggests a mass:

\centerline{$M_{BH}^{M_{BH}-\sigma} = 1.11^{+1.05}_{-0.54}\times 10^8 M_{\odot}$~~.}

Equations \ref{eq:M_L} and \ref{eq:M_sigma}, despite the fact that they are derived from the same sample by \cite{2013ARA&A..51..511K}, differ by almost an order of magnitude in terms of $M_{BH}$. Adopting $M_{BH}^{M_{BH}-\sigma}$, as a more robust result and calculating the magnitude that the bulge \emph{should have} to justify this mass, we find a minimum difference of:

\centerline{$\delta M_{K_S} \sim$ $1.7$ mag}

\noindent
or, in other words, NGC3718 should be at least $\sim$ $4.8$ times brighter. 

But what can cause such a disagreement between the two scaling relations, in the case of NGC3718? A first approach could be that, this is not a bulge but a pseudo-bulge\footnote{A pseudo-bulge is a central high density stellar region similar to a classical bulge, but with properties closer to disk-like objects (e.g. supported mainly from rotation rather than from pressure due to random motions of stars). Their formation mechanisms are believed to be different than the ones that formed classical bulges.}, so these equations do not apply \citep{2011Natur.469..374K}. NGC3718 is currently classified as an SB(s)a pec, meaning that it has a \emph{bar} and, according to \cite{2004ARA&A..42..603K}, barred galaxies preferably have pseudo-bulges.

On the contrary, several authors \citep[e.g][]{2004A&A...415...27P,2005A&A...442..479K,2009AJ....137.3976S} have studied the gas dynamics and have successfully fitted tilted rings on NGC3718. They all agree that the gas orbits are nearly edge-on near the center. \cite{2009AJ....137.3976S} suggest that the classification as a barred galaxy is misleading. They see the apparent nuclear bar, as a projection effect of the prominent dust lane of NGC3718, which is the result of looking through a disk/ring of dusty gas on polar orbit around the, nearly face-on, stellar disk. They also report that the latter must be substantially free of cool gas, since they do not detect any HI emission on the plane of rotation of the stellar disc, a characteristic of lenticular galaxies. Additionally, the L1.9 spectral classification \citep{1997ApJS..112..391H}, means that broad line $H_{\alpha}$ emission is detected, suggesting that our line of sight provides a direct view towards the nucleus of the host galaxy which, consequently, has to be closer to face-on. \cite{2009AJ....137.3976S} conclude that NGC3718 is a Polar Ring Galaxy, a rare class of objects, often hosting a lenticular galaxy, which as an early-type object, shares many kinematic and other properties with elliptical galaxies. Our data also favor this picture, since we do not see any traces of a nuclear bar. The elliptical region around the nucleus ($H-K_S$ map of figure \ref{fig:NIR_maps_1}), looks more like an almost face-on disk rather than a bar. 

Moreover, NGC3718 does not fulfill at least three of the criteria of \cite{2004ARA&A..42..603K}, for classifying a bulge as a pseudo-bulge, namely: 
\begin{enumerate}
\item Following \cite{2013ApJ...769L...5K}, we calculate the Faber-Jackson correlation for coreless and core ellipticals respectively, using $L_V = 1.20 \times 10^{10} L_{\odot}$, namely:

\begin{equation}
\ {L_{\mathrm{V}} \over 10^{11} L_{\odot}} = (0.67 \pm 0.10) \left({\sigma \over 250 km s^{-1}}\right)^{3.74\pm0.21} \rightarrow \sigma \sim 158  km s^{-1}
%\{L_\mathrm{V} \over 10^{11} L_{\odot}} = (0.67 \pm 0.10) \left({\sigma \over 250 km s^{-1}}\right)^{3.74\pm0.21} \rightarrow \sigma \sim 158  km s^{-1}
\label{eq:FB_coreless}
\end{equation}

\begin{equation}
\ {L_{\mathrm{V}} \over 10^{11} L_{\odot}} = (0.79 \pm 0.11) \left({\sigma \over 250 km s^{-1}}\right)^{8.33\pm1.24} \rightarrow \sigma \sim 199  km s^{-1}
%\ {L_{\mathrm{V}} \over 10^{11} L_{\odot}} = (0.79 \pm 0.11) \left({\sigma \over 250 km s^{-1}}\right)^{8.33\pm1.24} \rightarrow \sigma \sim 199  km s^{-1}
\label{eq:FB_core}
\end{equation}

NGC3718's kinematic properties \citep{2002A&A...393L..89J} imply the presence of a \emph{random motion supported bulge} and its velocity dispersion \citep{2009ApJS..183....1H} seems to be in good agreement with both versions of the Faber-Jackson correlation (given their large scatter), although a closer match to that of a coreless elliptical (equation \ref{eq:FB_coreless}). In any case, the observed $\sigma$ is not much smaller than the $\sigma$ predicted by these relations, which would be indicative of a rotationally supported pseudo-bulge. As pointed out by \cite{2013ARA&A..51..511K}, ``\emph{Classical bulges are essentially equivalent to coreless ellipticals}'', a picture consistent with the classification of NGC3718 as a lenticular galaxy. Considering also, 

\item that we do not detect any signs of a nuclear bar and 

\item that there is no sign of significant star formation\footnote{We support the view of \cite{2009AJ....137.3976S}. We examine GALEX near and far UV images and we find that, NGC3718's UV flux is $\sim1$ order of magnitude less than the flux of the starburst galaxy NGC4449, which indicates some amount of star formation, but at the same time, this is not high enough to be characterized as ``starburst''.} \citep{2009AJ....137.3976S}, 
\end{enumerate}
allowing us to assume that NGC3718 behaves as an elliptical galaxy and, therefore, it should follow the host galaxy - SMBH mass scaling relations. But why do the $M_\mathrm{BH} - M_{K_S,bulge}$ and $M_\mathrm{BH} - \sigma$ relations point towards different $M_{BH}$ ?

\subsection{Light ``deficit'' or $\sigma$ ``surplus''?}
\label{sec:missing_light}

Given the fact that equations \ref{eq:M_L} and \ref{eq:M_sigma} are calibrated against the same galaxy sample by \cite{2013ARA&A..51..511K}, we can safely assume that they should, more or less, agree. As previously noted (section \ref{sec:bulge}), we consider $\sigma$ to be more accurately determined than $M_{K_S,bulge}$, since, as it is measured from two different linewidths and the Faber-Jackson relation, shows surprisingly good consistency.

\begin{figure}[h]
\centering
\includegraphics[width=\columnwidth]{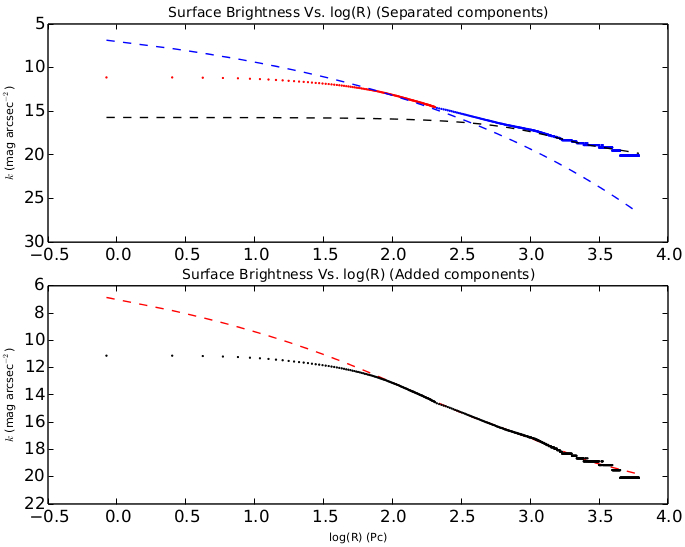}
\includegraphics[width=\columnwidth]{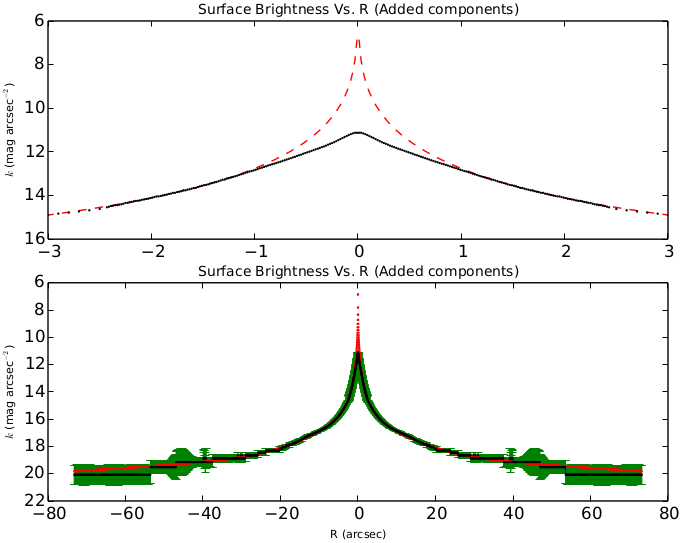}
\caption{First figure: Surface brightness in the $K_S$ band versus log(R) in pc. The blue dashed line represents the inwardly extrapolated Sersic profile of the bulge, while the black dashed line represents the dual exponential disk profile. The blue points represent the 2MASS $K_S$ data, while red points represent our SUBARU $K_S$ data. Second figure: Same as the first figure, but with the different components added. Here the red dashed line represents the fitted model whereas the black points represent the complete joined data set. Third figure: Surface brightness in the $K_S$ band versus R (in arcsec) of the innermost $\sim3$ arcsec. The red dashed line represents the fitted model, while black points represent (up to $\sim2.5$ arcsec) our SUBARU $K_S$ data. Fourth figure: Same as the third figure, but plotted for the complete data range ($\sim70$ arcsec). The green lines represent the uncertainties of the data points.}
\label{fig:surface_brightness}
\end{figure}

Our first choice therefore, is to examine the robustness of $M_{K_S,bulge}$. In order to better understand our light distribution, we produce the surface brightness profile of NGC3718, from a joined data set consisting of our SUBARU\footnote{Sample in the inner $\sim2.5$ arcsec.} and 2MASS\footnote{Sample between $\sim2.5$ to $\sim70$ arcsec.} $K_S$ data, against the radius. For this, we use IRAF's task \emph{ellipse} once again, which returns an azimuthally averaged surface brightness profile, as shown in figure \ref{fig:surface_brightness}. Many authors \citep[e.g.][]{2009ApJS..182..216K} argue that, fitting a Sersic profile to larger radii of an elliptical galaxy and then extrapolating it inwardly, is an accurate way to describe the surface brightness profile. 

Following this scheme, we fit our combined SUBARU and 2MASS $K_S$ data set of NGC3718. Our best fit consists of a Sersic profile for the bulge with an index $n = 4.85$, combined with a dual exponential disk component with scale lengths $R^{disk_{1}}_{S}\sim 7$ arcsec and $R^{disk_{2}}_{S}\sim 60$ arcsec respectively, as presented in figure \ref{fig:surface_brightness}. We choose the lower limit of our fitting range ($\sim$ $50$ arcsec), based on the uncertainties of the 2MASS data points (at $\sim 50$ arcsec the flux is $\sim$ $2\sigma$ above the background). The upper limit is then determined through $\chi^{2}$ minimization for that given lower limit, in the range between the (candidate) upper limit and $\sim 20$ arcsec, in order to minimize the fit with respect to the higher signal-to-noise data points (at $\sim 20$ arcsec the flux is $\sim$ $6\sigma$ above the background). Our best fit has a $\chi^{2}$ $\sim2$, for a data range between $\sim1.5 - 50$ arcsec ($\sim$ $500$ data points). Our modeled bulge magnitude estimation is $M_{K_S,bulge} \sim$ $-22.85$ magnitudes ($m_{K_S,bulge} \sim$ $8.35$). When put into equation \ref{eq:M_L}, this ``recovered'' magnitude, suggests an $M_\mathrm{BH}$ of the order of:

\centerline{$M_{BH}^{M_{BH}-M^{Sersic}_{K_S,bulge}} = 1.18^{+1.17}_{-0.59}\times 10^8 M_{\odot}$~~,}

which, surprisingly enough, covers the gap between the $M_\mathrm{BH} - L$ and $M_\mathrm{BH} - \sigma$ relations (section \ref{sec:bulge}), almost perfectly.

A similar result, was previously noted by \cite{2006AJ....131.1236D}, who used 2MASS data of NGC3718 along with the bulge/disk decomposition algorithm GALFIT, in order to separate the bulge from the disk and obtain magnitudes for both. Their best fit consists of a Sersic bulge profile of index $n = 5.5$ with $M^{Dong}_{K_S,bulge} \sim$ $-22.75$ magnitudes ($m^{Dong}_{K_S,bulge} \sim$ $8.45$) and a disk with scale length $R^{disk}_{S}\sim$ $60$ arcsec. This magnitude estimation is almost identical to ours. In this case, equation \ref{eq:M_L} then gives:

\centerline{$M_{BH}^{M_{BH}-M^{Sersic}_{K_S,bulge}} = 1.05^{+1.05}_{-0.53}\times 10^8 M_{\odot}$~~.}

What both of these fits reveal is very interesting, since, instead of a Sersic profile near the center, we get a \emph{plateau}. The fit, describes the surface brightness profile very well at radii $\geq 1-1.5$ arcsec, but it ``fails'' near the center, where the model suggests that an amount of ``missing light'' is present, seen as the difference between the fitted and the observed curves. \cite{2009ApJS..182..216K} suggest that a surface brightness profile of this kind, is a characteristic seen in core galaxies.

In short, a ``core'' is an elliptical galaxy whose measured surface brightness profile shows a break towards a shallower light profile when compared to an inward extrapolation of its outer Sersic fit, revealing an amount of light ``deficit''. The opposite stands for coreless galaxies. These are objects whose measured surface brightness profiles show a break towards a steeper light profile with respect to their, inwardly extrapolated, outer Sersic fit, revealing an amount of light ``surplus''.  

So, as our analysis suggests, the light ``deficit'' is, most likely, the source of the disagreement between the $M_\mathrm{BH} - L$ and $M_\mathrm{BH} - \sigma$ relations\footnote{We have to note at this point, that we agree with \cite{2008ApJ...679..156H} on the ``danger'' of doing three component fits. Our fit is indeed very sensitive to the adopted lower and upper limits. This is the reason why we resort in trusting the statistical robustness of our sample and set statistical, yet as physical as possible, constrains. The fact that we confirm, more or less, the fit of \cite{2006AJ....131.1236D} indicates some consistency. Moreover, we know that light is ``missing'' from the central region, since the published aperture photometry values for NGC3718 on NED are closer to our SUBARU measurement than to the ``recovered'' magnitudes. We can not, however, rule out the case of a bit different Sersic index with a small ``extra light'' component near the position of the transition from the disk to the bulge dominated component (log(R) $\sim2-2.5$ pc in figure \ref{fig:surface_brightness}). Such a transition would be very difficult, if not impossible, to be distinguished in this case due to the presence of the disk. Provided that NGC3718 is a merger remnant, the low star formation activity is expected to have created an amount of post merging stars, that could, in principal, have resembled an ``extra light'' component. We think that this would not change dramatically our overall view, especially regarding the amount of the ``missing light'' which seems to be quite dominant.}$^{,}$\footnote{The sensitivity issue on the adopted upper and lower limits, is also mentioned by \cite{2002A&A...393L..89J} and they decide not to make such a fit. Our SUBARU data, however, resolve the innermost $\sim2.5$ arcsec of NGC3718 much better, providing us with a richer data sample to work with.}. As discussed by several authors \citep[e.g][]{2009ApJS..182..216K,2013ARA&A..51..511K,2008ApJ...678..780G}, such a light ``deficit''\footnote{Observed up to now mostly on bright giant ellipticals.} could be indicative of co-evolution driven by a combination of \emph{SMBH binaries} followed by \emph{SMBH recoils}. The light ``deficit'', can be retrieved by the difference between the observed and the expected brightness (figure \ref{fig:surface_brightness}). In our case, it translates to an $M^{NGC3718}_{def} \sim 2.2 \times 10^{9} M_{\odot}$ and a ${M_{def} \over M_{BH}} \sim 19$. When put onto the $M_{def}$ versus $M_{BH}$ plot of \cite{2013ARA&A..51..511K} (their figure 30), it falls on the left of and away from the main distribution ($M_{def} \sim5 M_{BH}$), but well within the observed upper limit of $M_{def} \sim50 M_{BH}$. Moreover, if NGC3718 is viewed as an advanced merger \citep{2002A&A...388..407C}, then it falls much closer to the main distribution when compared to NGC4486B, the only other ongoing merger on this plot, with a ${M^{NGC4486B}_{def} \over M_{BH}} \geq 50$. Additionally, for a real(istic) core, a mass ``deficit'' of $\leq1\%$ of the total galaxy mass is expected \citep[e.g.][]{2008ApJ...679..156H}. Using $M^{NGC3718}_{tot.} \sim 400 \times 10^{9} M_{\odot}$ from \cite{1985A&A...142..273S} and calculating the percentage, we get an $M^{NGC3718}_{def}\sim 0.5\% M^{NGC3718}_{tot.}$, which is well within the realistic limits. These facts, make NGC3718's relatively large $M_{def}$, appear as, at least, physically plausible.

Finally, our fit is consistent with previous studies. \cite{2002A&A...388..407C} and \cite{2002A&A...393L..89J} studied 2MASS and kinematic data of NGC3718, for which they find elliptical-like kinematics mixed with spiral-like photometry. \cite{2005A&A...437...69B}, attributes the ``transitional'' nature of these objects to result from mergers with mass ratios $\geq (3-4.5):1$, remnants which closely resemble the properties of lenticular galaxies. The fact that we and \cite{2006AJ....131.1236D}, see both the exponential disk and the random motion supported bulge in the surface brightness profile means that NGC3718 should be classified as a lenticular galaxy which, consequently, strongly implies that NGC3718 is a merger remnant. 

The mass ``deficit'' and the indications for a merger, are compatible with the evidence we have so far, that NGC3718 might host an actual SMBH recoil. We discuss further this subject in the discussion (section \ref{sec:discussion}). For now, we adopt the average value of the equations \ref{eq:M_L} (ours and \cite{2006AJ....131.1236D}) and \ref{eq:M_sigma} for the rest of our calculations, as a good estimation for the $M_{BH}$, namely:

\centerline{$M_{BH}^{NGC3718} = 1.11^{+1.09}_{-0.55}\times 10^8 M_{\odot}$~~.}

\subsection{Core or coreless?}
\label{sec:classification}

As we have seen up to now, NGC3718 appears to show properties that belong to both core and coreless ellipticals. For example, its velocity dispersion fits more to that of a coreless galaxy, whereas the presence of ``missing light'' is a characteristic usually seen in core galaxies. It is reasonable, therefore, to investigate further similarities and differences with either of these categories. The classification criteria of the E-E dichotomy\footnote{Elliptical galaxies form two distinct categories, each one defined by specific fundamental properties such as core or coreless, boxy or disky etc.}, have been discussed by many authors \cite[e.g][]{2007MNRAS.379..401E,2005ApJ...621..673T,2007MNRAS.379..418C} and the most updated classification criteria for core and coreless elliptical galaxies, are presented by \cite{2009ASPC..419...87K}.

Supporting a classification as a coreless object is the visual magnitude $M_{V}^{NGC3718} = -20.73$ mag, which is fainter by almost an order of magnitude than the $M_{V} \sim -21.5$ mag, the ``border''\footnote{The E-E dichotomy ``border'' is calculated for $H_0=$ $72$ km s$^{-1}$. For this value the absolute magnitude of NGC3718 becomes $M_{V}^{NGC3718} = -20.60$ mag.} of the E-E dichotomy, lying on the bright end of the coreless side \citep[see figure 2 of ][]{2009ASPC..419...87K}. Additionally, the presence of weak/not dominant radio emission, as we show in the e-Merlin $6$ cm map (figure \ref{fig:emerlin}) and as measured by \cite{2007A&A...464..553K} and by us (see section \ref{sec:merlin}), which, along with NGC3718's kinematic properties \citep{2002A&A...393L..89J}, are characteristics usually seen in coreless objects.

\begin{figure}[h]
\centering
\includegraphics[width=\columnwidth]{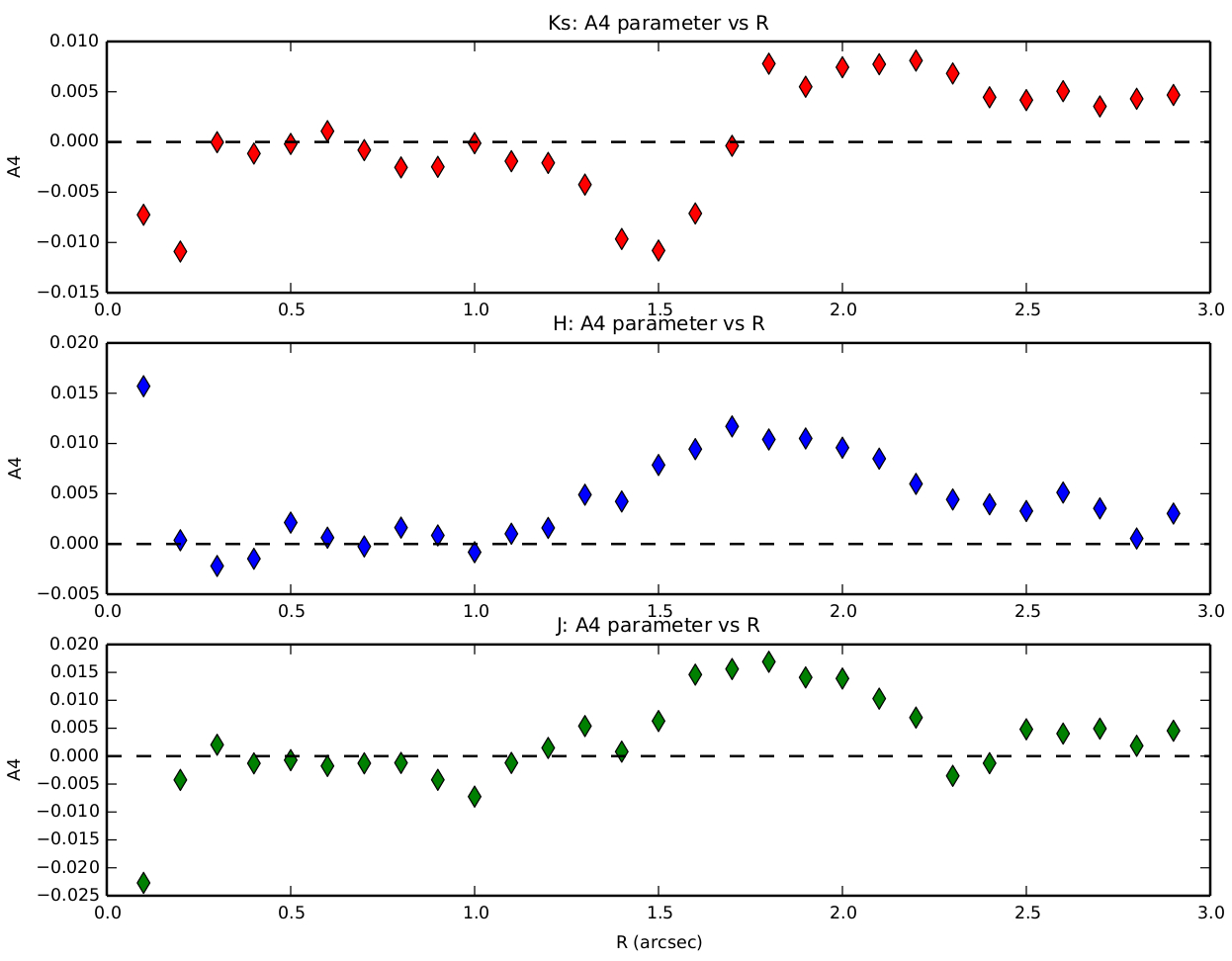}
\caption{$A4$ parameter versus radius for NGC3718. From top to bottom, A4 parameter for the $K_S$ (red/top), $H$ (blue/middle) and $J$ (green/bottom) bands respectively, versus R in arcsec. In all plots, the dashed black line represents the border between a ``disky'' and a ``boxy'' object.}
\label{fig:isophote}
\end{figure}

On the other hand, the most obvious observational evidence to support a classification as a core object, is the ``missing light'' that we observe in the surface brightness profile. The fact that the Sersic profile has an index of $n>4$ and the presence of moderate to low hard X-ray emission ($L^{SWIFT-BAT}_{X-rays} = 6.46 \times 10^{41}$ ergs sec$^{-1}$ from \cite{2010ApJS..186..378T}), may further strengthen a classification as a core object. 

Regarding the isophote contours shape of the images of figure \ref{fig:data}, we see the same general trend in all bands, as we present in figure \ref{fig:isophote}. The A4 parameter\footnote{A4 Fourier coefficient indicating the deviation from an ellipse. ``Boxy'' objects have a $A4<0$, whereas ``disky'' objects have a $A4>0$.} is mostly negative up to a radius of $\sim$ $1-1.5$ arcsec (essentially, the radius of the core we observe), consistent with a boxy (core) object, while further out it steadily raises above zero and drops again, to remain on the positive side, up to a radius of $\sim$ $3$ arcsec, consistent with a disky (coreless)  object. This could be indicative of an inside-out transformation of NGC3718 from a coreless to a core galaxy. The available classification criteria, are summarized in table \ref{tab:NGC3718_lit}.

\begin{table}[h]
\caption{E-E dichotomy parameters.}
\centering
\label{tab:NGC3718_lit}
\begin{tabular*}{\columnwidth}{@{\extracolsep{\fill}}cccc} \hline \hline
Criteria & core & coreless & Reference\\\hline
Visual Magnitude &  & $\checkmark$ & 1  \\
Surface Brightness profile& $\checkmark$ &   & 5 \\
Isophote Contours shape & $\checkmark$  & $\checkmark$ & 4 \\
Sersic index & $\checkmark$ &   & 5,6 \\
Radio emission &  & $\checkmark$ & 3,7,8,9 \\
X-ray emission & $\checkmark$ &  & 2 \\
Kinematics & & $\checkmark$ & 10 \\
\hline
\end{tabular*}
\tablefoot{
E-E dichotomy classification parameters for NGC3718.\\
}
\tablebib{
(1)~NED; (2) \citet{2010ApJS..186..378T}; (3) Figure \ref{fig:emerlin}; (4) Figure \ref{fig:isophote};
(5) Figure \ref{fig:surface_brightness}; (6) \cite{2006AJ....131.1236D}; (7) \cite{2004A&A...415...27P}; 
(8) \cite{2005A&A...442..479K}; (9) \cite{2007A&A...464..553K}; (10) \cite{2002A&A...393L..89J};.
}
\end{table}

The picture we get from the above discussion is that NGC3718 is most likely an intermediate (between core and coreless) galaxy, a fact that makes the attempt for a possible interpretation of the results more complex. It provides, however, an additional piece of evidence in favor of a view of NGC3718, as the ``smoking gun'' of a past/ongoing merger, in the sense that the E-E dichotomy refers to early-type objects, believed to have formed in mergers.

\section{Electromagnetic signatures of an SMBH recoil}
\label{sec:EM_sign}

So far, the observations imply that NGC3718 is a merger remnant, with a core in its surface brightness profile, combined with a spatial and a, potentially coincident, AGN offset, all of which are compatible with SMBH recoils. 

If this is indeed the case, according to \cite{2012AdAst2012E..14K}, we should expect to see some characteristic electromagnetic signatures. The most prominent of those, would be a velocity shift between the broad and the narrow component of a given line. Briefly, the idea is that a region of the order of size of the Broad Line Region (BLR) should be gravitationally bound to the recoiled SMBH, resulting in a relative motion of the BLR with respect to the, much larger, Narrow Line Region (NLR). \cite{1997ApJS..112..391H} fit a narrow and a broad component to the $H{\mathrm{\alpha}}$ line and they find that the centroid of the broad $H{\mathrm{\alpha}}$ is redshifted with respect to its narrow counterpart by $430$ km s$^{-1}$ (their figure 10e). If we also account for systematic offsets of the order of $50 - 100$ km s$^{-1}$, between broad and narrow line components \citep{1997iagn.book.....P}, we adopt a kick velocity of the order of:

\centerline{$v_{kick}\sim 355$ km s$^{-1}$~~.}

Based on the $v_{kick}$ and $M_{BH}^{NGC3718}$ estimations, we can calculate the gravitationally bound radius, according to the relation \citep{2006MNRAS.367.1746M,2012AdAst2012E..14K}:

\begin{equation}
R_{bound} = \left({G \times M_{BH} \over {v_{kick}}^2}\right) \approx 0.4 \times \left({M_{BH} \over 10^{8} M_\odot}\right)\left({v_{kick} \over 10^{3} km s^{-1}}\right)^{-2} pc
\label{eq:GBA}
\end{equation}

\noindent
which results to an $R_{bound} \sim 3.53$ pc. Since the gravitationally bound radius is inversely proportional to the $v_{kick}$\footnote{This velocity shift represents the projected kick velocity, so it may only serve as a lower limit of the true kick velocity and this should be accounted for, in all subsequent calculations.} and due to the large uncertainties of the $M_{BH}^{NGC3718}$, this should be viewed as an indicative, yet relatively uncertain, estimation. With this in mind, we can estimate the size of the BLR, using the relation from \cite{2007MNRAS.374..691Z}: 

\begin{equation}
R_{BLR} = q \times \left({G \times M_{BH} \over {FWHM_{H{\mathrm{\alpha}},Broad}}^{2}}\right) 
\label{eq:BLR}
\end{equation}

\noindent
which, depending on the adopted value of q, gives an $R_{BLR} \sim 0.06 - 0.12$ pc ($q = 0.75 - 1.33$), using $FWHM_{H{\mathrm{\alpha}},Broad} \approx 2350$ km s$^{-1}$ for NGC3718 from \cite{1997ApJS..112..315H}. One last quantity to estimate, is the sphere of influence of the SMBH, given by the relation:

\begin{equation}
R_{infl.} = \left({G \times M_{BH} \over \sigma^{2}}\right)
\label{eq:SOI}
\end{equation}

\noindent
which equals to $R_{infl.} \sim 19.1$ pc for the values we have adopted up to now in this paper. The latter is $\sim 5$ times smaller than the core radius, i.e. $\sim 100$ pc. 

These values are useful, in order to better understand the nuclear dynamics of the system. Specifically, the $R_{BLR}$ and the $R_{bound}$, strongly suggest that a BLR of the predicted size would likely be, entirely, gravitationally bound to a recoiled SMBH. 

For our final calculation, we use the images of figure \ref{fig:bulges} and we measure values below the resolution limit. The fact that these images have not been extensively tested for their scientific accuracy, makes them useful only for rough (order of magnitude) estimations (see section \ref{sec:evaluating the centering process}). 

With this in mind, the presence of a Hyper Compact Stellar System (HCSS) is expected around an SMBH recoil. According to \cite{2012AdAst2012E..14K}, an HCSS should be found within an $R_{bound}$ and its total luminosity should be of the order of the luminosity of a globular cluster. We attempt to measure this, by measuring the $K_S$ band fluxes of figures \ref{fig:bulges} and \ref{fig:data} at the position of the offset red blob, within an $R_{bound}$ and taking the difference. These two images differ only by the presence of the offset red blob in that region. The rest of the emission through this column, should be a good approximation of the ``foreground'' and ``background'' emitting sources, respectively, with respect to the center. This luminosity difference is:

\centerline{$L_{HCSS} \sim 3.5 \times 10^{4} L_{\odot}$~~,}

\noindent
whereas this converted to mass (see section \ref{sec:intro}), yields a:

\centerline{$M_{HCSS} \sim 2.7 \times 10^{4} M_{\odot}$~~,}

\noindent
which is within the predicted luminosity and mass range for a globular cluster. 

These findings, further strengthen the possibility that an SMBH recoil is indeed present in NGC3718, since the observations appear to be compatible with key aspects of the current theoretical view around SMBH recoils \citep[e.g.][]{2012AdAst2012E..14K,2011MNRAS.414.3656S}.

\section{Discussion}
\label{sec:discussion}

Summarizing the results, the observations suggest that NGC3718 appears to have, a spatial offset which, potentially coincides, with a radio emission offset, a relative redshift between its broad and narrow $H{\mathrm{\alpha}}$ lines and the presence of a core in its surface brightness profile, all compatible with the presence of an SMBH recoil. Moreover, the strong indications that NGC3718 is a merger remnant including, spiral-like photometry mixed with elliptical-like kinematics and properties that belong to both core and coreless ellipticals, fulfill the prerequisite for an SMBH recoil, in the sense that merging is the current theoretical mechanism for the formation of an SMBH binary that can lead to an SMBH recoil. Some of these indications, however, could also be viewed to apparently contradict each other. Thus, a very important question is the following: ``Is there a theoretical framework which could incorporate all the, potentially contradictory, characteristics, into one physically consistent scheme?''.

\subsection{Formation of cores and galaxy evolution}
Attempting to give a decent answer to this question, requires us to briefly review what we know, up to now, about the two relevant sub-classes of the elliptical galaxies family. In recent years, it has become widely accepted that, within the $\Lambda$$CDM$ cosmology, structures grow hierarchically in the Cosmos \citep[e.g.][]{1978MNRAS.183..341W}. Within this picture, classical bulges and elliptical galaxies are formed in major mergers, resulting in the common properties observed in these objects. Moreover, elliptical galaxies define a trend of increasing dissipation with decreasing mass \citep{2009ASPC..419...87K}. This means that later-type, fainter ellipticals, being ``1st stage'' merger products of gas-rich spiral galaxies, have more gas to dissipate through ``wet'' merging \citep{2009ApJS..181..135H}, whereas earlier type giant ellipticals, being products of the subsequent merging of gas-poor fainter ellipticals, were formed in ``dry'' mergers \citep{2009ApJS..181..486H}. 

An immediate effect of ``wet'' mergers is that older stars (formed in the progenitors before merging), relax violently at larger radii (giving rise to a Sersic law profile), while new stars are formed mainly in a central intense starburst due to dissipation. As the cold gas experiences tidal torques due to the violence of a major merger, is channeled towards the central region, triggering SMBH growth, SMBH feedback and an intense nuclear starburst \citep{2008ApJ...679..156H}. The latter forms the ``extra light'' component in these objects, which is observationally recognizable as a break towards a steeper inner surface brightness profile, when compared to an inward extrapolation of the outer profile's Sersic fit \citep[e.g.][]{2009ApJS..181..135H,2009ApJS..181..486H,2013ARA&A..51..511K}. During this process the SMBH accretes up to the point that the energy feedback output from the AGN is large enough ($\sim 1 \%$ of a, near the Eddington limit, AGN's energy output is coupled to the infalling gas) to blow away the residual gas, stopping the BH growth and quenching the star formation in the central region \citep[e.g.][]{2005RSPTA.363..667O}. SMBH recoils are still expected to occur, since merging is a prerequisite of the process of elliptical galaxies formation. The presence of gas, however, helps the SMBH recoil to return to the center more quickly \citep[e.g.][]{2011MNRAS.412.2154B,2011MNRAS.414.3656S}, stopping any core scouring and replacing the ``missing light'' with newly formed stars in the central starburst \citep{2013ARA&A..51..511K}, which in turn, consumes the larger fraction of the available gas \citep{2008ApJ...679..156H}. During this process, the AGN operates in quasar mode.

However, ``dry'' mergers, as a product of merging of fainter gas-poor ``extra light'' ellipticals, lack the intense central starburst. The old violently relaxed stellar populations of the progenitors will again violently relax at larger radii in the newly formed system, whereas the ``extra light'' components will end up near the center and be preserved, in the form of a more compact stellar distribution in the central region \citep{2009ApJS..181..486H}. Although tidal forces will channel the (little) remaining gas towards the center, the newly formed giant elliptical is massive enough to hold onto enough amounts of X-ray emitting gas, which ensures that any cold gas reaching the nuclear region will be heated up, preventing any star formation \citep{2007MNRAS.382.1481N}. The AGN in this case operates in the, so called, maintenance mode, which further helps the gas to remain hot, making the nuclear star formation even more difficult and, eventually, keeping the merger ``dry'' \citep{2013ARA&A..51..511K}. Due to the absence of gas, SMBH recoils, are able to ``stay on duty'' for a longer time, acting mainly on a spacial range of the order of the preserved ``extra light'' fossil in the newly formed remnant, scouring the core and creating what appears as a ``missing light'' or, in other words, a break towards a shallower inner surface brightness profile, when compared to an inward extrapolation of the outer profile's Sersic fit.

\subsection{Formation of polar ring galaxies}
As we extensively discuss in section \ref{sec:bulge}, the observations support NGC3718's classification as a polar ring galaxy, suggested by \cite{2009AJ....137.3976S}. But how are polar ring galaxies formed? 

There are two candidate formation mechanisms: The merging scenario, proposed by \cite{1997ApJ...490L..37B,1998ApJ...499..635B}, suggesting formation through a head-on collision between two orthogonally placed spiral galaxies, the intruder and the victim (left panel of their figure 1). After the merging the intruder becomes the host galaxy (often an S0), whereas the victim becomes the polar ring. The second is the accretion scenario. Supported by a number of authors \citep[e.g.][]{1997A&A...325..933R}, it explains polar ring galaxy formation through gas accretion from a donor galaxy, as a result of tidal gravitational interactions due to a close encounter. Subsequent merging of the two galaxies is not necessary for the formation of a ring, but it is not physically prohibited. 

\cite{2003A&A...401..817B} simulated both of these ideas, using the same numerical model. They find that both mechanisms successfully reproduce many of the observed properties of polar ring galaxies such as, a) early-type host galaxies, b) stable, nearly polar, ring structures (some of their runs are evolved for $8-10$ Gyr with a typical formation time for a polar ring galaxy of $\sim 3$ Gyr) and c) warps/spiral arm like structures when these objects are viewed from specific lines of sight (edge-on/face-on for the ring/host galaxy respectively). All of these properties are observed to be present in NGC3718, by a number of independent studies \citep[e.g.][]{2004A&A...415...27P,2005A&A...442..479K,2009AJ....137.3976S}.

A few properties, however, differentiate the two proposed scenarios. Firstly, in the merging scenario the host galaxy has to be substantially free of gas (which is the case of NGC3718 according to \cite{2009AJ....137.3976S}), whereas in the case of the accretion scenario the host galaxy can also be a gas-rich object. Secondly, the merging scenario is expected to form a faint, spherical, diffuse stellar halo, consisting of the stars of the victim galaxy and surrounding both the host galaxy and the polar ring. Such a faint diffuse stellar component can be seen in figure \ref{fig:SDSS_comb} of NGC3718, nearly engulfing the edge-on ring and the host galaxy. However, in the accretion scenario, instead of such a component, a donor galaxy has to be identified in a relatively close proximity, which in this case could be the close neighboring galaxy NGC3729 or a similar galaxy that has already been cannibalized by NGC3718. Lastly, the merging scenario appears to be more consistent, in (inefficiently) transferring gas towards the central region ($\sim 10-25 \%$ in most of their runs), than the accretion scenario. \cite{2003A&A...401..817B} quote that, the gas infall is not as large and as general in the accretion scenario as it is in the merging scenario. They conclude that the accretion scenario is the most likely formation mechanism, at least, for the majority of the cases without, however, ruling out the merging one based on some physical explanation, but rather through statistical argumentation, leaving space for it to actually occur in Nature.

To the extent that a subsequent merging, after a possible accretion of gas into polar orbit, is not a common process, our analysis tends to favor the merging scenario for the formation of NGC3718. The reason for this is mainly the observational evidence we present throughout this work for the presence of an SMBH recoil, which requires a merger to occur. Moreover, the low star formation activity (also seen by \cite{2009AJ....137.3976S}) and the detection of rapidly rotating molecular gas within the central $\sim 700$ pc seen by \cite{2005A&A...442..479K} (through their position-velocity diagrams), also favor the merging scenario, since the gas inflow predicted by this appears to be consistent with these observations. Finally, if the dust component traces the kinematics of the gas (a reasonable assumption since, gas and dust are mixed in the apparent polar ring), then the contribution from hot dust to the $K_S$ band emission that we detect (see section \ref{sec:light_decomp}), serves as an additional indirect tracer for the aforementioned gas inflow. 

The accretion scenario can also, in principal, explain these phenomena. Provided that a subsequent merging is a common outcome of this scenario, then this could be viewed as a merging event of different orbital configurations. So, when we refer to a ``merging event'', we do mean either of the aforementioned polar ring galaxy formation scenarios, as long as the merging of SMBHs is ``guarantied'', in order to explain the indicated SMBH recoil and the observed core.

\subsection{Putting the pieces together}
The inefficient transport of gas towards the nucleus, however, might be the key in providing a robust physical explanation for the core that we see in the surface brightness profile of figure \ref{fig:surface_brightness}. A working scheme could be the following:

If NGC3718 is viewed as a gas-rich\footnote{\cite{2009AJ....137.3976S} measures $M_{HI} \sim 8 \times 10^{9} M_{\odot}$, a value $\sim2$ times larger than the Milky Way, while \cite{2005A&A...442..479K} measures $\sim 2.4 \times 10^{8} M_{\odot}$ of molecular gas through CO observations, values that can characterize NGC3718 as a gas-rich galaxy.} merger remnant, one would expect to see signs of the intense nuclear starburst in the form of an ``extra light'' component, as described by, e.g. \cite{2013ARA&A..51..511K,2008ApJ...679..156H,2009ApJS..181..135H}, instead of the observed core. The suggested absence of sufficient amounts of nuclear gas, however, lead naturally to the limited star formation observed. This, combined with the fact that in the absence of sufficient amounts of gas, the time scale for which an SMBH recoil ``stays on duty'' is prolonged \citep{2011MNRAS.414.3656S}, provides a physically consistent framework for the formation of the observed core. Put more simply, the gas exists in the galaxy but it is ``locked up'' on (the stable) polar orbit \citep{2003A&A...401..817B,1997ApJ...490L..37B,1998ApJ...499..635B}, a fact that reduces the  available gas that can flow towards the center, triggering the necessary star formation that could ``fill the gap'' created by SMBH scouring.

Additionally, the behavior of the A4 parameter (figure \ref{fig:isophote}), indicative of an inside-out transformation from a disky to a boxy object, fits naturally to this framework. Boxyness is present in objects that lack the intense central starburst, either because of limited amounts of nuclear cold gas and/or star formation quenching due to AGN feedback \citep{2013ARA&A..51..511K}. Such an environment seems ideal for an SMBH recoil to excavate a core. This process appears to be connected with the isophote contours shape behavior in the case of NGC3718, since the transition from boxy to disky, occurs roughly in the spatial scales of the observed core.

In this framework, the presence of energetic X-ray emission ($14-195$ Kev) reported by \cite{2010ApJS..186..378T}, also fills a gap. Provided that it is associated with AGN feedback, it makes the star formation quenching mechanism applicable in this case. The very good agreement between the velocity dispersion measured by the gas emission and the stellar absorption features \citep{2009ApJS..183....1H} ($\sigma_{gas} \over \sigma_{\star}$ is $\sim 1$ see section \ref{sec:bulge}), could be considered as the observational consequence of this process. 

Any apparent contradictions originate from the fact that, all these processes do not take place in a ``red and dead'' giant elliptical (for which they were originally proposed) but, circumstantially, due to the limited gas inflow attributed to the nature of the dynamics of this (rare) merger case, in a gas-rich system.

The last of NGC3718's properties that we have to incorporate into this scheme, is its spectral classification. NGC3718 is classified as a LINER (low-ionization nuclear emission line region) galaxy \citep{1997ApJS..112..391H} and it is also part of the NUGA sources, a survey aimed at the study of nearby low-luminosity active galactic nuclei (LLAGNs) \citep{2003ASPC..290..423G}. This is also consistent with the inefficient transport of gas towards the center, in the sense that it affects the accretion rate by reducing the available gas reservoir. The latter is implied by the fact that, NGC3718 appears to be a sub-Eddington system with ${L_{bol} \over L_{edd}} \sim 10^{-4}-10^{-5}$ \citep{2007A&A...464..553K}, indicating that the accretion rate onto the central engine is far from being efficient.

\subsection{On the SMBH recoil}
The last question within the scope of this study, is, perhaps, the most prevalent one, albeit a very difficult one to answer: ``Is this indeed a case of a true SMBH recoil?''. An attempt to qualitatively answer this question requires us to further discuss the compatibility of an SMBH recoil with other observational findings for NGC3718. 

The near-the-center position of the offset red blob, allows space for two different scenarios: It could either represent the beginning of an SMBH displacement, or, the ``capture'' of an ongoing SMBH recoil in one of its pericentric passages.

At first, if the age of $\sim2-3$ Gyr of NGC3718's gas disc, suggested by \cite{2009AJ....137.3976S}, traces roughly the age of the merger, then it is in excellent agreement with the results of the simulations of \cite{2003A&A...401..817B}, where a stable polar ring and a warp is formed within $\sim2-3$ Gyr, from mergers with mass ratios of $\sim(1-4):1$. \cite{2011MNRAS.412.2154B} simulated recoils in galaxy mergers and they find that, for mergers with mass ratios $\sim(1-2):1$ and gas content of $10-30\%$, the coalescence of the SMBHs often occurs at $\sim2$ Gyr. For such systems and for kick velocities of $v_{kick} \leq 0.7v_{esc}$, the trajectories of the kicked SMBHs are often confined within $\sim1$ kpc with setback times of the order of $\sim1$ Gyr. So, time wise, it seems plausible that either a new or an ongoing SMBH recoil exists in NGC3718 and that it would be observable.

Moreover, assuming that our current view for the formation of cores is attributed mainly to SMBH binaries and the subsequent recoils \citep[e.g.][]{2013ARA&A..51..511K}, then the rather large $M_{def}\sim19 M_{BH}$ (see section \ref{sec:missing_light}), needs to be addressed. Such a large $M_{def}$ could be partly attributed to accumulated errors from i.e. uncertainties in the fit, uncertainties in the $M \over L$ relation and/or due to the large intrinsic scatter of the scaling relations. Such large mass deficits, however, could also imply that pre-existing cores in the progenitor galaxies have been inherited to NGC3718. \cite{2008ApJ...678..780G} simulated kicks and the mass deficits that these can produce. They find that a combination of a pre coalesced binary \citep{2006ApJ...648..976M} followed by a subsequent recoil, can excavate as much as $\sim5 M_{BH}$ per merger. The fact that the light profiles are (more or less) preserved in dissipationless mergers \citep[e.g.][]{2009ApJS..181..486H}, is what leads several authors to the conclusion that, large mass deficits could be indicative of different core excavation events following sequential mergers. A large $M_{def}$ therefore, seems possible to co-exist, with either a new or an ongoing SMBH recoil.

NGC3718's radio emission, however, may help us draw a line between a new and an ongoing SMBH recoil. To the extent that NGC3718 is indeed an X-shape source, it could be the host of a recently recoiled SMBH that has undergone a reorientation of its jet \citep{2012ApJ...746..176L}. In this case, the inheritance of a core of the observed size ($\sim200$ pc across) is considered necessary, since a just recently recoiled SMBH would not have the time to excavate such a large (both spatially and in terms of $M_{def}$) core \citep{2006ApJ...648..976M}. If, on the contrary, the secondary extensions on the radio map are not attributed to a second jet-fossil, then, an ongoing SMBH recoil which accretes during one of its pericentric passages \citep{2011MNRAS.412.2154B}, could produce a single bipolar jet structure. In this case, the inheritance of a core of the observed size is not necessary (but, partly, not prohibited either). So, in principal, a core could be excavated in ``one go'', provided that the $M_{def}$ is considerably overestimated due to the aforementioned uncertainties. 

To conclude this work, we have presented much independent evidence, which, along with our measurements on the SUBARU and e-Merlin data, suggest that it is reasonable to treat NGC3718 as a very good candidate host galaxy of an SMBH recoil. Therefore, an interesting question arises: ``Is the presence of SMBH recoils in mergers, a standard process in galaxy evolution?''. If, as the observations suggest, NGC3718 is indeed a polar ring galaxy, and polar ring galaxies make up $\sim5\%$ of the lenticular and early type galaxies in the local universe \citep{1990AJ....100.1489W}, then, the detection of an SMBH recoil in such a rare object may, by itself, suggest an affirmative answer to the above question. In any case, this is as far as our data allow us to go in interpretation terms. Our limited knowledge on the exact behavior of recoiled SMBHs, our incomplete understanding of the precise mechanism(s) that are responsible for the formation of cores and the limited resolution of the current observations, allow space only for general qualitative, and therefore highly speculative, interpretations. Certainly though, better quality future observations combined with higher resolution simulations, are necessary, in order to answer the above questions in a more robust, quantitative way. In all, NGC3718 turns out to be an object full of wonderful surprises with, possibly, a lot more that are yet to be revealed.

\section{Summary}
\label{sec:summary}

We have extensively studied NGC3718, using NIR SUBARU and 6cm e-Merlin radio data as well as previously published results. Our findings are summarized as follows: 

\begin{enumerate}

\item Our NIR color maps do not show any signs of a large scale constant color gradient, indicating that our view towards the nucleus of NGC3718 is unaffected by extinction.

\item An offset red blob is detected in our NIR color maps, being displaced by $\sim4.25$ pc from the center of the underlying stellar bulge. The radio emission appears to originate from a position closer to the offset red blob than to the center of the stellar bulge.

\item A light decomposition reveals a contribution (up to $\sim50\%$) from a hot (up to $\sim1000$K) dust component to the incoming light from the central $\sim0.5$ arcsec, indicative of the presence of gas in that region.

\item An extended elongated structure $\sim1$ arcsec across, is detected in the e-Merlin radio map, probably indicative of a small scale bipolar jet. A second smaller one $\sim0.6$ arcsec across, lies almost perpendicular to the first one, suggesting a possible X-shape radio source. However, these extended structures may also contain contribution from SNR (Supernova Remnant) related non-thermal emission.

\item A disagreement between the $M_\mathrm{BH} - L$ and $M_\mathrm{BH} - \sigma$ scaling relations when NGC3718's measured values are used and the shape of the surface brightness profile, reveal an amount of missing light in the form of a scoured core of $\sim200$ pc across. This light ``deficit'' translated to mass, reflects an $M_{def}\sim19M_{BH}$.

\item NGC3718 shows mixed characteristics in two ways: On one hand, it has spiral-like photometry (seen in the surface brightness profile as a dual disk component) combined with elliptical-like kinematics (seen in the surface brightness profile as a bulge component with Sersic index $n>4$). Furthermore, NGC3718 shows characteristics that belong to both, fainter, more rotationally supported, disky, coreless ellipticals, as well as to giant, less rotationally supported, boxy, core ellipticals. These, strongly imply that NGC3718 is the result of a merger.

\item Finally, NGC3718 appears to be the host of an SMBH recoil. Evidence for that include an offset NIR red blob, with a, potentially coincident, offset radio emission, the presence of a core in the surface brightness profile, a relative redshift between the broad and narrow $H{\mathrm{\alpha}}$ emission lines and the indication for the presence of a Hyper Compact Stellar System surrounding the offset NIR red blob.

\end{enumerate}

\begin{acknowledgements}

We thank Prof. David Merritt, for the enlightening discussions that helped us forge a better understanding of the nuclear dynamics. We would also like to thank the anonymous referees, for the fruitful comments that helped clarify the description of the methods we used in this work. The SUBARU telescope operation team as well as the HiCIAO instrument team for generous support. MERLIN is a National Facility operated by the University of Manchester at Jodrell Bank Observatory on behalf of STFC. This work was supported by the Max Planck Society and the University of Cologne through the International Max Planck Research School (IMPRS) for Astronomy and Astrophysics as well as in part by the Deutsche Forschungsgemeinschaft (DFG) via grant SFB $956$.  We had fruitful discussions with members of the European Union funded COST Action MP$0905$: Black Holes in a violent Universe and the COST Action MP$1104$: Polarization as a tool to study the Solar System and beyond.

\end{acknowledgements}

\bibliographystyle{aa}
\bibliography{NGC3718} 

\begin{appendix}
\section{On the influence of the dust lane}
\label{sec:app_A}

Appendix \ref{sec:app_A}, provides additional evidence to the ones discussed in section \ref{sec:nir maps}, regarding the potential influence of the dust lane on our view towards NGC3178's nucleus. 

\begin{figure}[h]
\centering
\includegraphics[width=\columnwidth]{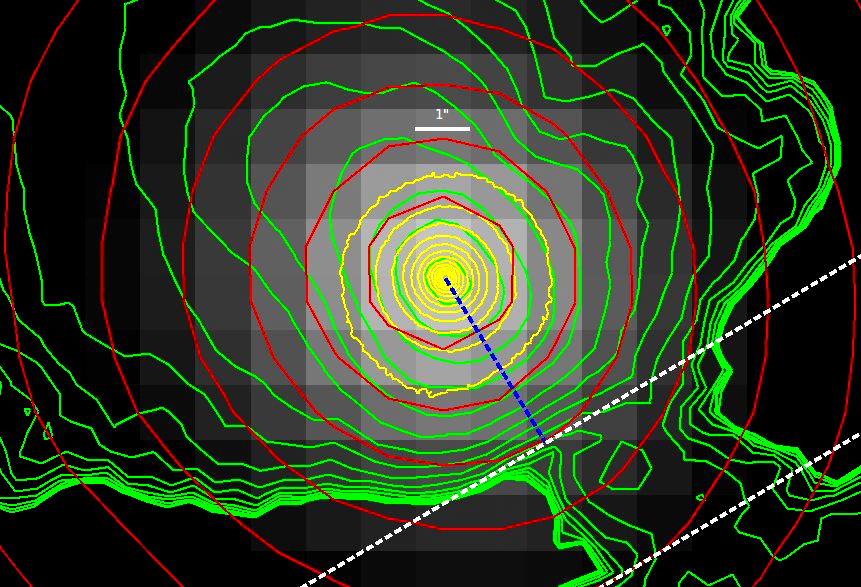}
\caption{2MASS $K_S$ band image of NGC3718. Over-plotted can be seen: 1) 2MASS $K_S$ band contours (red), 2) SDSS $r$ band ($\sim0.6~\mu$m) contours (green), 3) SUBARU $K_S$ band contours (yellow) and 4) distance (blue dashed line length $\sim3.5$ arcsec) of the optical and NIR peak flux positions to the approximate beginning of the dust lane (outlined with the white dashed box).}
\label{fig:dust_1}
\end{figure}

For this reason we present figure \ref{fig:dust_1}, where the 2MASS $K_S$ band image of NGC3718 is displayed. What can be noted in figure \ref{fig:dust_1}, is the fact that the 2MASS $K_S$ band contours (in red) appear to be, almost entirely, unaffected by the presence of the dust lane. They appear to be significantly rounder and better defined, especially when compared to the SDSS $r$ band contours (in green), which are severely deformed by the presence of foreground dust in the region. Additionally, our SUBARU $K_S$ band contours (in yellow) appear to be more of a smaller scale inward extrapolation of the 2MASS $K_S$ band contours (as expected), rather than of the visible light, heavily, distorted contours. Moreover, the innermost $\sim4$ arcsec of our SUBARU $K_S$ band, are located $\sim1.5$ arcsec away from the region where the contours deformation in the optical becomes catastrophic, which naturally, coincides approximately with the projected beginning of the dust lane (white dashed box).

We consider as safe, therefore, to treat the central region of NGC3718 in the NIR, as being unaffected by the presence of the dust lane, when we attempt to interpret our NIR color maps.

\end{appendix}

\begin{appendix}
\section{On the $z$-$J$ bands alignment}
\label{sec:app_B}

Appendix \ref{sec:app_B}, focuses on the validity of the coordinates calibration method used on the SUBARU data, described in section \ref{sec:uncertainties}.

\begin{figure}[h]
\centering
\includegraphics[width=\columnwidth]{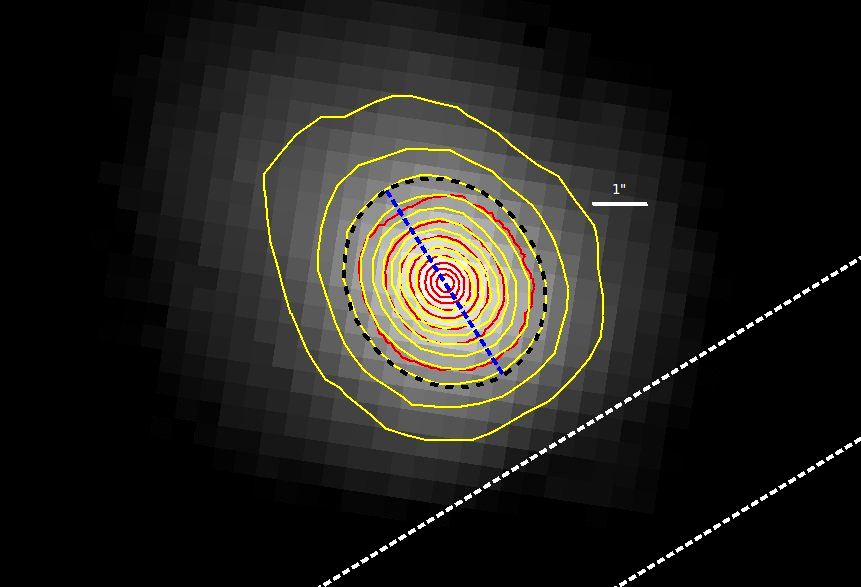}
\caption{SDSS $z$ band image of NGC3718. Over-plotted can be seen: 1) The SDSS $z$ band contours (yellow) and 2) the SUBARU $J$ band contours (red). The black dashed ellipse and the blue dashed line show the symmetry of the SDSS $z$ band contours within the central $\sim4$ arcsec of NGC3718.}
\label{fig:dust_2}
\end{figure}

For this reason we present figure \ref{fig:dust_2}, where the SDSS $z$ band image of NGC3718 is displayed. In this image, the black dashed ellipse has a major axis of $\sim4$ arcsec, while the blue dashed line indicates that this is, approximately, the radial distance where the SDSS $z$ band contours begin to deviate from being symmetric with respect to the peak flux positions. What can be seen, is that the SDSS $z$ band contours are well defined, relatively round and, more importantly, symmetric within this region. Especially when compared to the SDSS $r$ band contours (green contours in figure \ref{fig:dust_1}), they seem to be almost entirely unaffected by the presence of dust within the region of interest and we consider them to be suitable for use, for an accurate estimation of the photocenter of NGC3718 in this band. Moreover our $J$ band contours appear to be similar to the SDSS $z$ band contours, i.e. in terms of similar roundness. The $z$ band ellipticities range between $\sim0.13-0.15$, while in $J$ band they range between $\sim0.09-0.12$, in the radial interval $\sim0.5-1.5$ arcsec, indicating that, at least to a large extent, both bands appear to be similarly unaffected by the presence of foreground dust. 

We consider, therefore, that the aforementioned symmetry of the SDSS $z$ band contours in the central $\sim4$ arcsec of NGC3718, as well as their similarity to the $J$ band contours, safely allow us to apply the centering method described in section \ref{sec:centering}. The derived photocenters are used to align the SDSS $z$ band with our SUBARU $J$,$H$ and $K_S$ bands, in order to astrometrically calibrate them.

\end{appendix}

\begin{appendix}
\section{On the e-Merlin radio map position}
\label{sec:app_C}

\begin{figure}[h]
\centering
\includegraphics[width=\columnwidth]{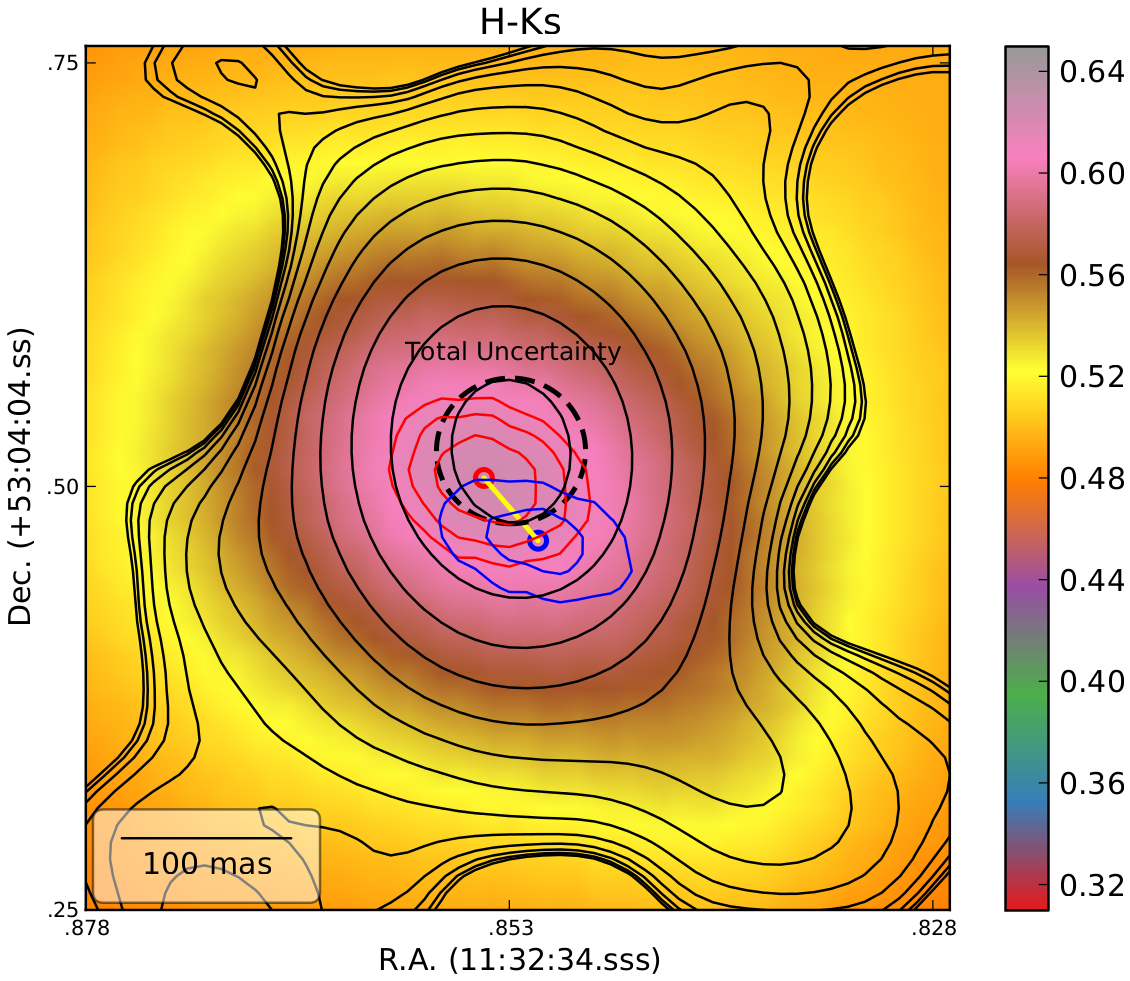}
\caption{Same as figure \ref{fig:H_Ks_emerlin}, but with the radio map position dictated by the old phase reference source position. This is the equivalent position to the one reported by \cite{2007A&A...464..553K}, namely R.A.: 11:32:34.8534 $\pm$ 0.0005, Dec.: +53:04:04.523 $\pm$ 0.004.}
\label{fig:H_Ks_emerlin_old_calib}
\end{figure}

\end{appendix}

\end{document}